\documentclass{aa} 
\let\tmp\newinsert
\let\newinsert\newbox
\usepackage{floatrow}
\let\newinsert\tmp
\usepackage{natbib}
\usepackage{graphicx}
\usepackage{txfonts}
\usepackage[normalem]{ulem}
\usepackage{xcolor}
\usepackage{etoolbox}
\usepackage{bm}
\usepackage{ulem}
\usepackage{amsmath}
\usepackage{multicol}
\usepackage{supertabular}
\usepackage{hyperref}

\makeatletter
\renewcommand*\aa@pageof{, page \thepage{} of \pageref*{LastPage}}
\def\ignore#1{}
\newcommand{\orcit}[1]{\protect\href{https://orcid.org/#1}{\protect\includegraphics[width=8pt]{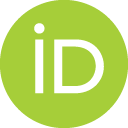}}}

\makeatletter
\renewcommand*\maketitle{%
  \thispagestyle{firstpage}
\begingroup
    \if@wideboxfn
    \setlength\bibindent{1.4\parindent}
    \else
    \setlength\bibindent{\parindent}
    \fi
    \renewcommand*\thefootnote{\@fnsymbol\c@footnote}%
    \renewcommand\@makefntext[1]{%
    \ifaa@longfn\hsize\textwidth\fi
    \noindent
    \hb@xt@\bibindent{\hss\@makefnmark\enspace}##1}
  \ifaa@twocolumn
  \begingroup
    \begin{aa@strip}
          \aa@maketitle
    \end{aa@strip}
    \@thanks	  	
  \endgroup
  \else
    \begingroup
      \let\thanks\footnote
      \aa@maketitle
    \endgroup
  \fi
\endgroup
  \setcounter{footnote}{0}%
}
\makeatother

\def\fdeg{\hbox{$^\circ$}}

\newcommand{\kms}{km.s$^{-1}$}

\newcommand{\ROHSA}{{\tt ROHSA}}
\newcommand{\HI}{\ifmmode \mathrm{\ion{H}{I}} \else \ion{H}{I} \fi}

\def\comments{true}
\ifx\comments\undefined
\def\QD#1{}
\def\CH#1{}
\def\RL#1{}
\def\JLV#1{}
\def\CB#1{}
\def\FA#1{}
\def\AM#1{}
\else
\def\QD#1{{\color{blue} \textsl{\small $^{[QD]}[$#1]}}}
\def\CH#1{{\color{red} \textsl{\small $^{[CH]}[$#1]}}}
\def\RL#1{{\color{green} \textsl{\small $^{[RL]}[$#1]}}}
\def\JLV#1{{\color{orange} \textsl{\small $^{[JLV]}[$#1]}}}
\def\CB#1{{\color{purple} \textsl{\small $^{[CB]}[$#1]}}}
\def\FA#1{{\color{brown} \textsl{\small $^{[FA]}[$#1]}}}
\def\AM#1{{\color{brown} \textsl{\small $^{[AM]}[$#1]}}}
\fi

\begin{document} 

   \title{Toward a 3D kinetic tomography of Taurus clouds: }

  \subtitle{II. A new automated technique and its validation\thanks{Table \ref{tab:points} is available in electronic form at the CDS via anonymous ftp to cdsarc.cds.unistra.fr (130.79.128.5) or via https://cdsarc.cds.unistra.fr/cgi-bin/qcat?J/A+A/ }}

   \author{Q. Duch\^ene\orcit{0000-0003-3224-5603}
          \inst{1}
    \and
        C. Hottier\orcit{0000-0002-3498-3944}\inst{1}
     \and
           R. Lallement\orcit{0000-0002-1284-875X}\inst{1}
       \and 
           J.L. Vergely\orcit{}\inst{2}
       \and
          C. Babusiaux\orcit{0000-0002-7631-348X}\inst{3}
           \and
           A. Marchal\orcit{0000-0002-5501-232X}\inst{4}
            \and
           F. Arenou\orcit{0000-0003-2837-3899}\inst{1}\thanks{Corresponding author: F. Arenou}
 }
   \institute{GEPI, Observatoire de Paris, Universit\'{e} PSL, CNRS, 5 Place Jules Janssen, 92190 Meudon, France\\
              \email{Frederic.Arenou@obspm.fr}
\and
   ACRI-ST, 260 route du Pin Montard, 06904, Sophia Antipolis, France       
         \and
             Univ. Grenoble Alpes, CNRS, IPAG, 38000 Grenoble, France
         \and
             Canadian Institute for Theoretical Astrophysics, University of Toronto, 60 St. George Street, Toronto, ON M5S 3H8, Canada            }

\date{Received 2023-01-23; accepted 2023-04-05}

\abstract
{Three-dimensional (3D) kinetic maps of the Milky Way interstellar medium are an essential tool in studies of its structure and of star formation. 
}
{
We aim to assign radial velocities to Galactic interstellar clouds now spatially localized based on starlight extinction and star distances from Gaia and stellar surveys.
}
{We developed an automated search for coherent projections on the sky of clouds isolated in 3D extinction density maps on the one hand, and regions responsible for CO radio emissions at specific Doppler shifts on the other hand. The discrete dust structures were obtained by application of the Fellwalker algorithm to a recent 3D extinction density map. 
For each extinction cloud, a technique using a narrow sliding spectral window moved along the contour-bounded CO spectrum and geometrical criteria was used to select the most likely velocity interval.}
{
We applied the new contour-based technique to the 3D extinction density distribution within the volume encompassing the Taurus, Auriga, Perseus and California molecular complexes. From the 45 clouds issued from the decomposition, 42 were assigned a velocity. The remaining structures correspond to very weak CO emission or extinction. We used the non-automated assignments of radial velocities to clouds of the same region presented in paper I \ignore{\citep{Ivanova21}}
and based on KI absorption spectra as a validation test. The new fully automated determinations were found in good agreement with these previous measurements.
}
{
Our results show that an automated search based on cloud contour morphology can be efficient and that this novel technique may be extended to wider regions of the Milky Way and at larger distance. We discuss its limitations and potential improvements after combination with other techniques.  
}


  
\titlerunning{3D kinetic tomography of the interstellar medium} 
\authorrunning{Duchêne et al.}

\maketitle
%

\section{Introduction}\label{intro}

Three-dimensional (3D) kinetic tomography of the interstellar medium (ISM), that is, the assignment of radial or 3D velocities to interstellar structures, is a multi-scale tool in star formation studies. On small scales and Milky Way ISM observations, it helps linking the extremely detailed multi-wavelength, multi-species emission measurements with localized structures and understanding the respective roles of self gravity, converging flows and turbulence, magnetic field and chains of reactions following the first star births. On a large scale, it allows the role of global dynamics and spiral arms to be distinguished from that of local structure and the local stellar population. Importantly, 3D kinetic maps of the ISM would be especially useful when used in conjunction with Gaia-based 3D kinetic maps of the Milky Way stars under development. The use of the Galactic rotation curve, although being convenient to locate distant Galactic interstellar clouds, is not very helpful in the aforementioned detailed studies due to limited realism, lack of detection of peculiar, therefore particularly interesting velocities, and is not applicable towards the Galactic center \citep{Wenger18,Peek22}. Fortunately, massive stellar surveys with new generation spectrographs, new detailed radio emission spectral data and, last but not least, parallaxes and photometric data from the Gaia mission, all brought during the last two decades, are currently bringing huge amounts of data that may enter 3D kinetic tomography computations. Combinations of astrometric and/or photometric distances with absorption or extinction measurements feed the construction of dust or gaseous species 3D  distributions, and, in turn, these distributions can be combined with emission spectra to get the 3D kinetic information.

Thanks to this favorable context, 3D kinetic tomography of the Galactic ISM was recently worked out in several ways. Each of the techniques has its own strengths and limitations. A detailed introduction to the kinetic tomography is given by \cite{Tchernyshyov17}, and additional descriptions and references are given in the articles cited below. A first approach is the combination of dust extinction data and radio emission spectra. \cite{Tchernyshyov17} developed a method using \HI\ and CO spectral data (i.e., position-position-velocity (PPV) cubes) on the one hand and cumulative reddening radial profiles derived from PansSTARRS and 2MASS on the other hand. Each spatial cell along a radial profile is assigned a Gaussian velocity distribution. The mean velocity is allowed to fluctuate around the value predicted by a radially varying rotation curve, the velocity dispersion is limited to a range of classical values and the velocity dispersion around this value has imposed limits. Conversion factors from CO equivalent width to H column and from H column to reddening are imposed. The adjustment of the model to the emission data is regularized to avoid strong variations among neighboring cells. The technique has the advantage of being applicable everywhere, but does not always allow to disentangle two (or more) clouds of similar dust opacity located at different distances and with different radial velocities, despite the regularization. A different method was developed by \cite{Zucker18} who extended the \cite{Green18} Bayesian technique for reddening profile determination from individual extinctions by fitting a linear combination of $^{12}$CO velocity slices to the major reddening steps corresponding to the densest structures in Perseus. The authors could assign radial velocities and distances to those structures in very precise way. The limitation here is due to the need for hypotheses about the absence of CO associated with the clouds in front of and beyond the considered region. 

A second category of kinetic tomography is based on massive amounts of stellar spectra and makes use of both interstellar absorption Doppler shifts and target distances. \cite{Tchernyshyov18} used Doppler shifts and strengths of a near-infrared diffuse interstellar band (DIB) extracted from SDSS/APOGEE stellar spectra in combination with photometric distances of the target stars to derive a large scale planar map of the radial velocity. The advantage of the method is the lack of ambiguities since one follows the parallel evolutions with distance of the DIB radial velocity and of the DIB strength. Potential limitations of this method may arise from the variability of the DIB to extinction ratio. However, according to \cite{Puspitarini15}, such variability should have weak consequences on the tomography. The authors used two different DIBs in the visible range and their measurements in a series of directions, and found very similar radial profiles of the strengths of both DIBs as well as of dust extinction. On the other hand, there are limitations in the achievable velocity resolution as a result of the non-negligible intrinsic width of the DIB. E.g., there is a limit of $\simeq$ 10 {\kms} in the case of the APOGEE DIB. Finally, the technique is still limited by the relatively low number of target stars possessing high spectral resolution, high signal spectra. A hybrid method, also based on absorption data, this time in combination with 3D extinction maps, was attempted by \cite{Ivanova21} (paper I). The authors used neutral potassium absorption data in a series of stars distributed within $\simeq$ 600\,pc in the anti-center area encompassing the Taurus, Auriga, Perseus, California clouds (hereafter mentioned as Taurus clouds) to derive in a non-automated way partial information on the velocity field. In this case, the kinematic information could benefit from the high spectral resolution and the choice of narrow and strong absorption lines (the 766.5 and 770.0\,nm KI lines) to ensure easy detection and Doppler shift precision. On the other hand, the very limited number of targets and their small spatial density led to a restricted study. A benefit of this study is its use as a validation test for the new automated technique described in this article. In all techniques based on absorption lines or bands, massive spectroscopic surveys with high sky coverage would be necessary to extend the 3D kinetic mapping. 

A third and completely different approach uses the fact that newly born stars share the motion of their parent interstellar clouds. \cite{Galli19} developed and applied clustering algorithms to identify groups of young stellar objects (YSOs) in Taurus star-forming areas and derived their distances and 3D motions based on VLBI astrometry, Gaia parallaxes and proper motions. They verified the similarity of the radial velocities of each cluster and the one of the associated molecular cloud, providing distance assignment to the clouds and newly born stars. The technique allows to reach high accuracy on distances and velocities, but is limited to star forming regions. In addition, infrared data are needed for the very young objects linked to their parent cloud due to the very strong dust opacity and extinction in the visible.  


In this work, we explored a different, automated technique. This novel method uses CO emission spectra and 3D dust extinction density maps. It is based on the resemblance between the projection of a cloud on the sky and the extent and shape of series of mono-kinetic CO emission maps. We applied this cloud contour-based technique to the volume encompassing the nearby anti-center clouds Taurus, Auriga, Perseus and California. The difficulty of performing a kinetic tomography in a region where dense clouds extended in distance and with overlapping 2D projections having a range of velocities relatively narrow (i.e., less than -15 to +15\,{\kms}) led us to choose these clouds as test cases to evaluate our method. This choice was also especially motivated by the availability of the aforementioned results recently obtained by \cite{Ivanova21}, allowing to perform a thorough validation. An additional reason for the choice of this region is its latitude range and the subsequent limited distance to the clouds (less than $\simeq$ 1\,kpc for $b<-3$ deg), which ensures their inclusion in the used 3D extinction density map.  Finally, as said above, the choice of these clouds is not guided by the search for an easy case. On the contrary, this region is very complex and has characteristics that make the kinetic tomography especially difficult. We know that in the particular case of two (or more) clouds at distributed distances and with identical projections on the sky, assignments of the two (or more) radial velocities measured in their directions are not possible, if these assignments are based on morphology only. Our study aims at evaluating the frequency of such cases of indeterminacy, or equivalently the applicability of a technique solely based on projection geometry to complex regions such as the Taurus area.

The paper is organized as follows.
Section~\ref{data} presents the data used as sources for our technical study. Section~\ref{codata} describes how the CO data are denoised by means of the {\tt ROHSA} code \citep{marchal19} and used to create a {\it restriction} mask necessary for the following step. Section~\ref{dustmaps} presents the decomposition of the masked dust distribution into distinct 3D structures. Sections~\ref{method} and \ref{results} detail the newly developed technique of velocity assignment to the dust structures and the results. Sections~\ref{validation} and \ref{perspectives} discuss the comparisons with other independent results, limitations and potential extensions of the method.



\section{Data}\label{data}

\subsection{CO data}
We made use of radio $^{12}$CO spectral cubes taken from the well known series of data assembled by \cite{Dame2001}. In the Taurus area the angular resolution is $\simeq$0.125 \fdeg and the CO emission spectrum is decomposed into channels of 1.3\,{\kms} each. The limits of the main region containing Taurus, Auriga, Perseus and California are $[150^\circ,185^\circ]$ in longitude and $[-30^\circ,0^\circ]$ in latitude. The characteristic noise for Taurus according to \cite{Dame2001} is $\simeq$ 0.25\,K.

This CO dataset was used at two different steps of the whole procedure and in two different ways. First, it was used to delimit regions of the sky in the direction of CO clouds (see section \ref{codata}). For this purpose, it was previously regularized and decomposed into discrete structures to avoid highly irregular contours of the resulting mask. Second, it was used in its initial state during the association phase described in section \ref{method}. 

As we will see, the angular and velocity resolutions of the CO data are suited for our study, given the minimum size of the dust clouds (see below). The main limitation here will come from the noise level.

\subsection{Extinction density maps}

We used a 3D distribution of extinction density based on work presented by \cite{Lallement22} and \cite{Vergely22}. The map is  computed through tomographic inversion of large amounts of extinction-distance pairs for stars distributed in both direction and distance, to produce the extinction per unit distance photons suffer when traveling at a given location in 3D space, a quantity proportional to the dust volume density at this point. The map is based on 35 millions of distance-extinction measurements based on Gaia parallaxes and Gaia plus 2MASS photometric data, augmented by 6 millions of measurements from published catalogs based on spectroscopy and photometry.  To overcome the strong under-determination aspect of the full 3D inversion (distribution in each point in space, based on a limited number of lines of sight), a regularization is done that takes the form of a spatial covariance kernel or minimum size of structures. In order to maximize the achieved spatial resolution (minimum kernel), an iterative, hierarchical method is used, under the form of a series of inversions at increasing resolution. In regions where the target space density is high, the kernel width is reduced,  while prior results are kept where the target density is scarce and does not allow to go further. For details about the inversion method and descriptions of the recent maps, readers can refer to \cite{Lallement19,Vergely22}. We have used the latest and most precise map for which the final (minimum) kernel width is 10\,pc. This resolution is achieved in a large fraction of the volume that is considered here. The map is discretized into voxels of $5\times5\times5$\,pc$^3$ and the unit of the 550\,nm extinction density is mag\,pc$^{-1}$. The minimum size of the inverted dust structures will be the main limiting factor of our study.

\begin{figure}
    \centering
    \includegraphics[width=0.49\textwidth]{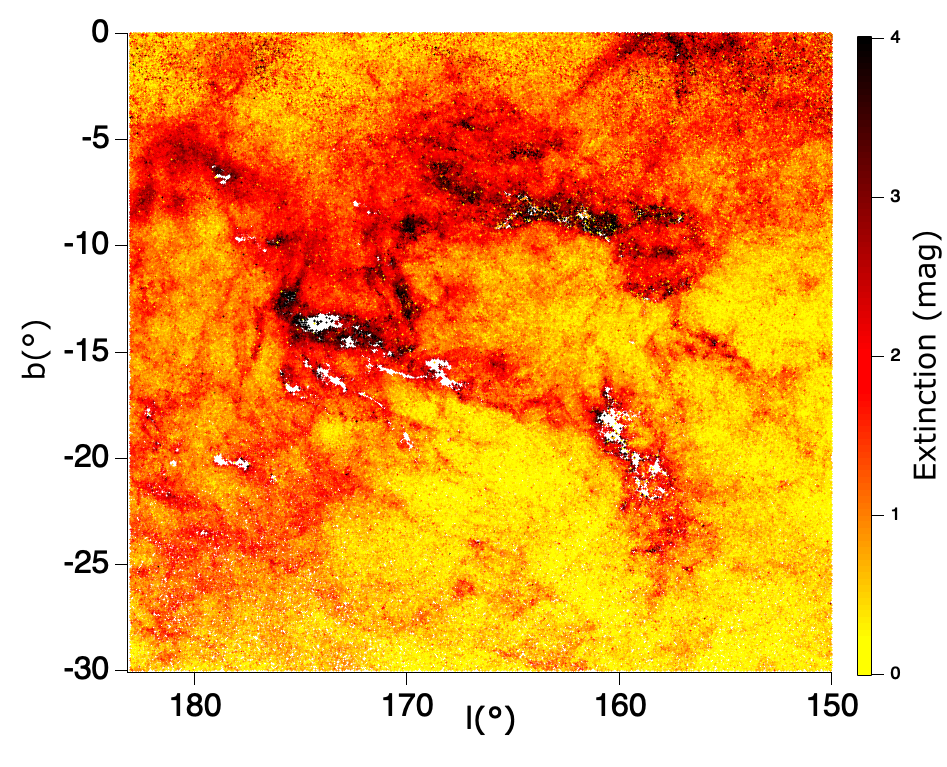}
     \caption{Target stars that have been used to build the 3D extinction density map in the Taurus area.  The locations of the stars are shown as dots colored according to their extinction (color scale at right). White areas correspond to high opacity, central regions within the main dust clouds (extinctions above $\simeq$ 4 to 5 mag) and mark the subsequent lack of targets with precise distances and photometry due to their faintness. The reconstructed dust distribution will reproduce the clouds but underestimate the corresponding high opacity of such cores.}
    \label{fig:targets_ext}
\end{figure}

In addition to the limitations on the spatial resolution associated with the limits on the target spatial density and the resulting local kernel, there is a second mapping limitation associated with lack of completeness in target brightness. As illustrated in Fig.~ \ref{fig:targets_ext}, there is a lack of target stars located behind the most opaque cloud cores in the input catalog of distance-extinction pairs, simply because foreground extinction makes them too faint to possess measurable parallaxes and/or accurate enough photometric or spectrometric data. As a result of the lack of strongly extincted targets, the reconstructed extinction density in the direction of the very dense cloud cores will be biased toward lower values than the actual ones, and this may affect dust to gas ratios estimates. However, the impact on our velocity assignment here is expected to be negligible, because we only use the clouds locations and their projected outer contours. 

\section{Decomposition of the 3D extinction density map into individual clouds}\label{dustmaps}

\subsection{CO-based restriction mask}\label{codata}

Our choice of using CO, a tracer of the dense molecular clouds, implies that the dust associated with clouds made of diffuse atomic gas only can not be part of an association with CO velocities. For this reason, we chose to eliminate as much as possible clouds that have no CO emissions to prevent unrealistic associations. To do so, we built and applied a  restriction mask to our dust map before its decomposition into clumps. We assigned a null extinction density to all locations in 3D space whose projections on the sky are within regions with null CO emission. In other words, we carved, or hollowed the 3D distribution. The mask does not eliminate atomic clouds located in the same direction as molecular clouds, but provides some useful limitation. 

%
%
%
In what follows, we describe the method used to build the mask. The reasons for our choice are detailed and illustrated in the appendix \ref{restriction_masks}.  First, we used the Gaussian decomposition algorithm \ROHSA\ \citep{marchal19} to denoise the CO data encompassing the neighboring anti-center clouds. 
In contrast to earlier applications dedicated to phase separation of 21\,cm data \citep[e.g., ][]{marchal-EN-2021,marchal_2021b} \ROHSA\ was used here for the spatial regularization it performs, which significantly reduces the noise. This regularization is a key step for the subsequent Fellwalker approach. The decomposition used in this work was obtained using $N=8$ Gaussian functions and the hyper-parameters $\lambda_{\bm{a}}=\lambda_{\bm{\mu}}=\lambda_{\bm{\sigma}}=10$ which controls the strength of the spatial regularization. As recommended in \cite{marchal19}, the number of Gaussian components and the amplitude of these three hyper-parameters were chosen empirically so that the solution converges towards a noise dominated residual. 

The second step was the decomposition of the resulting CO data model into discrete elements in the $(l,b,v)$ space, based on the Fellwalker code \citep{hottier21}. This algorithm determines areas of high flux by looking for each path with sufficient gradient coming to a local maximum. All paths leading to the same maximum form a clump. To do so, we tuned the parameters of the Fellwalker algorithm as follows: a minimum value to be considered part of a clump of 0.20\,K, which is the standard deviation of values on this map, a maximum size of jump of 3 voxels and a minimum depth of valley between two neighboring peaks of 1\,K. Here, voxels measure $0.125^\circ \times$ $0.125^\circ \times$ 1.3\,{\kms}.

The mask was finally defined as the full area of the projection on the sky of the totality of the retrieved elements resulting from the decomposition. Fig.~\ref{fig:contours_clumps} shows the shape of this restriction mask. After being superimposed on a CO contour at 0.11\,K.{\kms}, we could check that it encompasses all the bright CO emitting regions and goes further out over a small distance where CO starts to be difficult to see.  We note that, independently of the construction of the mask, this Fellwalker decomposition in $(l,b,v)$ domains can not be used for the velocity assignment, because the algorithm does not allow two distinct clouds to share the same triplet $(l,b,v)$, a situation commonly encountered in areas such as the one considered here. 

\subsection{Decomposition}

\begin{figure}
     \includegraphics[width=1\columnwidth]{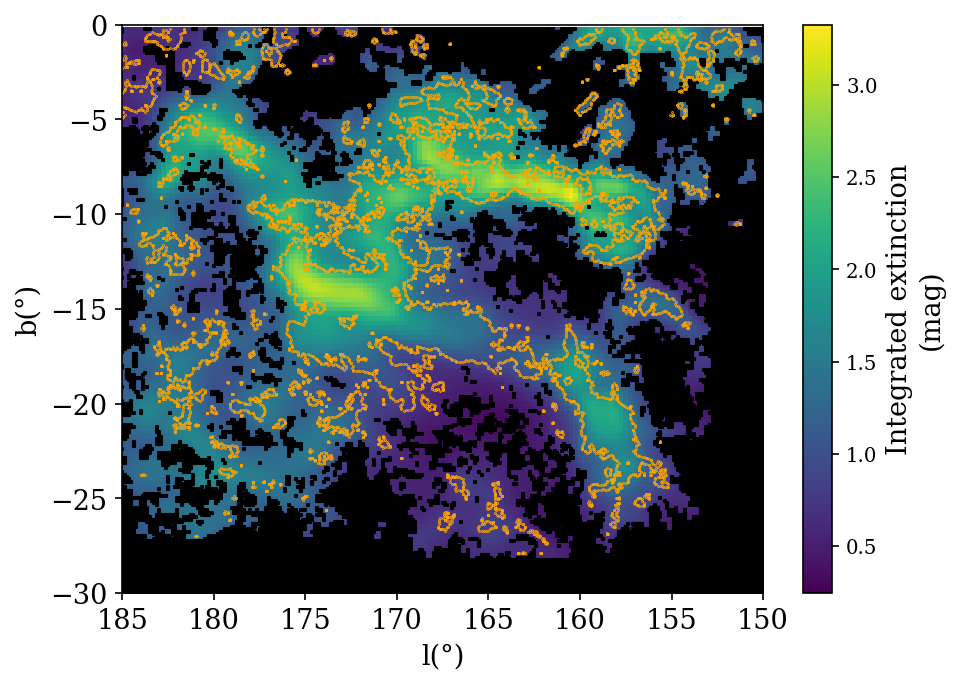}
     \includegraphics[width=1\columnwidth]{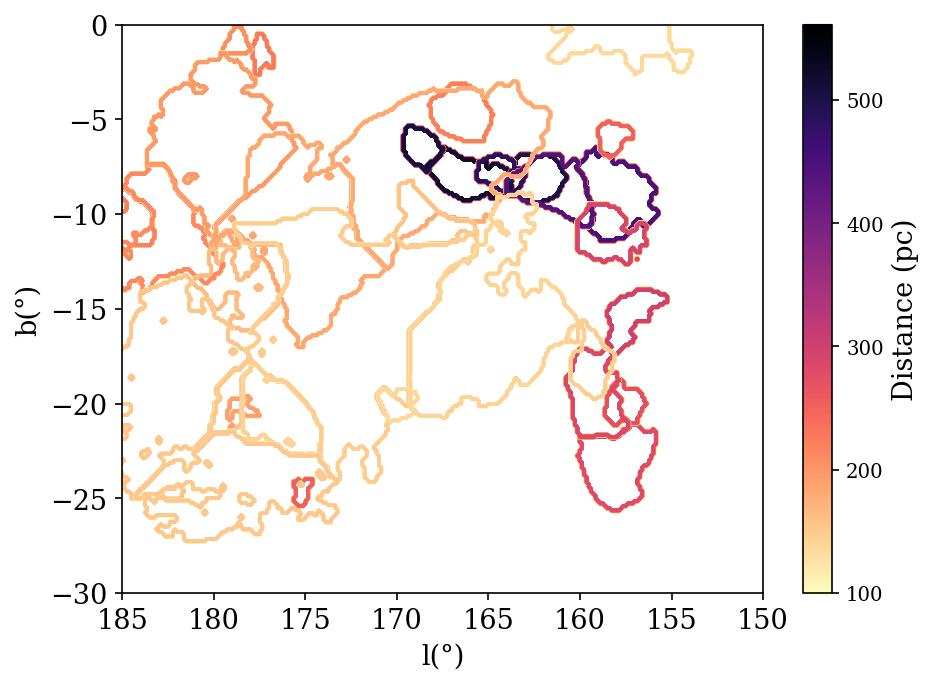}
    \caption{{\bf Top}: Restriction mask determined from the CO spectra (see section \ref{codata}) and applied to the 3D extinction map before its Fellwalker decomposition step. Masked areas are in black and superimposed on an image of the cumulative extinction integrated between 0 and 1\,kpc within the inverted 3D extinction density distribution. The yellow contour corresponds to an integrated CO intensity of 0.11\,K.{\kms}. The unmasked areas encompass all regions where CO is present, including areas of very weak intensity close to the measurement uncertainty level, except very close to the Plane due to the limited size of the 3D map (see text). {\bf Bottom}: Projection in the same area of the discrete extinction clouds issued from the Fellwalker decomposition. Cloud contours are colored according to their mean distance from the Sun (right scale).
    }
    \label{fig:contours_clumps}
\end{figure}

As in \cite{hottier21}, we choose to use Fellwalker algorithm \citep{Berry15} to extract from the 3D dust extinction density map individual structures. As explained above, we started with the modified extinction density map, in which non-null extinction voxels were restricted to non-masked directions (Fig \ref{fig:contours_clumps}). 
The gradient-based Fellwalker algorithm has the advantage of preventing the clustering from being dominated by radially elongated structures (the so-called {\it fingers of God}). Such features result from uncertainties on target stars distances and are common in 3D maps obtained by inversion. The radial gradients inside the {\it fingers of God} are almost flat, therefore they can be filtered with the minimum gradient parameter. In the case of the extinction map used in this study, based on a hierarchical process and on the selection of accurate distances, the {\it fingers of God} are quite limited. However, the cavities we created by application of the CO mask produce artificially elongated structures which are mimicking elongated {\it fingers of God}. We tuned the parameters of the algorithm as follows: a threshold minimum value to be considered part of a clump of 5.76 mmag, a maximum size of jump of 2 voxels, a minimal gradient of 2.5 mmag/voxel, a clump outline size of 1 voxel, and a minimum depth of valley between two neighbor peaks of half a standard deviation of the cube. We have chosen this last parameter in order to limit the merging step to the strict minimum.  Doing so, we made the algorithm give us smaller clumps. We did this in order to obtain more precision about velocity variations with our method and because we were confident that it is robust enough to work well despite this over-division of our map. With this set of parameters we sought to extract only the densest clumps and keep them as tight as possible, because we intended to compare their angular positions to CO emission which has a better angular resolution than the extinction map. The threshold extinction value was computed from the threshold for the emission mask and the mean dust-to-CO ratio in Taurus, Perseus and California from \cite{LewisEtAl2022a}.
Fig.~\ref{fig:contours_clumps} shows the position of clumps in longitude-latitude coordinates. We can easily discern the three  main features: Taurus  and Auriga in foreground, the more distant Perseus at $l\in[155^\circ,160^\circ]$ and $b\in[-25^\circ,-17^\circ]$ and the most distant California  at $l\in[155^\circ,170^\circ]$ and $b\in[-11^\circ,-4^\circ]$, as well as some other isolated clumps. The figure also illustrates two types of decomposition.  On the one hand, some clouds have overlapping projections, fully disconnected contours and lie at significantly different distances. On the other hand, there are also pairs of neighboring clouds at the same distance and adjacent to each other like two neighboring pieces of a puzzle, that clearly result from the division of a single, large volume into two distinct, smaller entities. This latter situation is due to our choice of the {\it valley} depth parameter. This will have an impact on our velocity assignment algorithm, as we will see below.

\section{Description of the velocity assignment method}\label{method}

Our goal is to assign a mean CO velocity to each 3D structure previously extracted from the extinction map, based on the CO spectral cube. The velocity assignment is guided by the following considerations: if a cloud, (hereafter central cloud or CC), either isolated or part of a contiguous more extended structure (called EC, see below) is moving at a radial velocity $V_i$, then two conditions should be fulfilled. First, this velocity should correspond to an intensity level above the noise in the CO spectrum. Second, and most importantly, the angular area S($V_i$) made by grouping all directions characterized by detectable CO emission at this specific velocity $V_i$ should bear some resemblance with the CC or EC projection. Following these requirements, a determination of the series of velocities measured over the CC angular area is firstly done, followed by their evaluation and ranking based on geometrical criteria favoring a CC source. For each individual cloud CC, we performed the following operations.

\subsection{Definitions}


\paragraph{Definition of a median emission profile.}

We selected the set of lines of sight of the CO grid encompassed by the CC projected contours, and for each of them we extracted from the CO spectral cube the individual CO temperature brightness profile as a function of LSR velocity. Based on this series of spectra, we computed for each velocity channel the median value over all points of the grid and combined the results to build a median profile. This profile contains all velocity components present within at least half of the CC angular area, associated to this central clump CC or to clouds in the foreground or in the background of this central cloud. This also includes components at the same velocity formed in spatially distinct clouds, in that case the additional contributions of these clouds may result in a total presence of the velocity in more than half of the CC angular area. The noise level for the median profile is  assumed to be 0.25\,K, the average value estimated by \cite{Dame2001} in the Taurus region, divided by the square root of the number of lines of sight in the central clump. \\ 

\paragraph{Definition of the associated extended structure.}

As we chose to configure Fellwalker to have the smaller clumps possible, we need to treat clouds resulting from the decomposition that are sub-parts of a wide cloud by taking account of contiguous clumps where necessary,  when we come to use a presence criterion (see below) to assign a radial velocity. This is why we considered not only the central clump CC but also its close neighbors, if they exist. These close neighbors were defined as those clumps whose projections fully or partially overlap the CC area and whose barycenter is located within 25\,pc from the barycenter of the central clump. This 25\,pc limit is determined to be greater than the characteristic size of clumps, and smaller than the distance between two separate clouds. The combination of the central clump and those neighbors is the EC structure. In case of an isolated cloud, EC reduces to CC.\\    

\paragraph{Definition of a test extended area.}

We considered a wide angular region called WA encompassing the contours of the central clump and of four times the central clump's area. To do so, we started with the smallest longitude-latitude rectangle containing the CC and extended it by adding the same number of pixels in latitude and in longitude until its area reached four times the CC area. The size of this region results from a compromise between the importance we want to give to the CC with respect to its angular environment, which favors a small area of the WA region, and the necessity to avoid cases in which WA would appear too small because the EC covers an extended area around the CC. Our choice of four times the CC area does not totally prevent the second case to happen, but it guarantees that the CC is the main feature in the region.\\

\paragraph{Definition of an {\it presence} ratio for a velocity interval.}

The velocity interval (or window) was defined by its central velocity $w_i$ and its width measured in CO velocity channels.  The criteria for the selection of the window and its central velocity $w_i$ were as follows: for each line of sight, we extracted the mean of the CO emission integrated along the window and compared it to the mean of the cloud median profile defined above. If the mean in the line of sight was greater than in the median profile and if the median profile at the central velocity had a brightness greater than the noise level of the median profile, we considered that the velocity window was present along this line of sight (the line of sight is selected). We then repeated this computation for each line of sight within the wide area WA, and computed the number of selected lines of sight interior to the clump's neighborhood (EC) contours $N_{ECi}$ on the one hand, and interior to the whole wide area WA $N_{WAi}$ on the other hand, as well as the ratio of these two numbers, that is the ratio of presence inside EC over the presence in WA, hereafter called {\it presence} ratio $P_{ECi}$:\\ 

\begin{align}
\label{eq:PEC}
    &&P_{ECi}=\frac{N_{ECi}}{N_{WAi}}
\end{align}


This selection process allows in principle to discriminate between the velocity component(s) associated with the EC, for which one expects a grouping of the selected lines of sight within the EC contours, and the velocities associated with closer or more distant clouds, for which the majority of the selected lines of sight are not expected to be linked with the cloud projected area. In the same way, we defined an  {\it presence} ratio specific to the central cloud alone $P_{CCi}$:\\

\begin{align}
\label{eq:PCC}
    &&P_{CCi} = \frac{N_{CCi}}{N_{WAi}}
\end{align}

\


\subsection{Velocity selection processes}

\paragraph{Initial selection using a sliding window along the median profile.}

We moved a sliding window in LSR velocity along the CC median profile. The area emitting at velocities comprised within the window interval was used to estimate the {\it presence} ratio defined above. For this first step of velocity selection, we used a window of three velocity channels (for a total of 3.9\,{\kms}) width, and a wider one of five velocity channels (for 6.5\,{\kms}) in order to cover different widths of components. We retained those window center velocities with the two highest ratios, and not located on adjacent velocity channels. More precisely, we first retained the center velocity with the highest ratio, then we looked at the second best that is not adjacent to the first. If no second best passed our criterion, we simply retained the first as prior. If no velocity passed our criterion, we rejected any velocity assignation for this clump. These first velocity intervals are too wide and too uncertain to be considered as components on their own, but they will serve as the best priors for the next step, a multi-Gaussian fit of the median profile.\\

 \paragraph{Velocity selections based on a Gaussian fit of the median profile.}

\begin{figure}
    \centering
    \includegraphics[width=1\columnwidth, height=6cm]{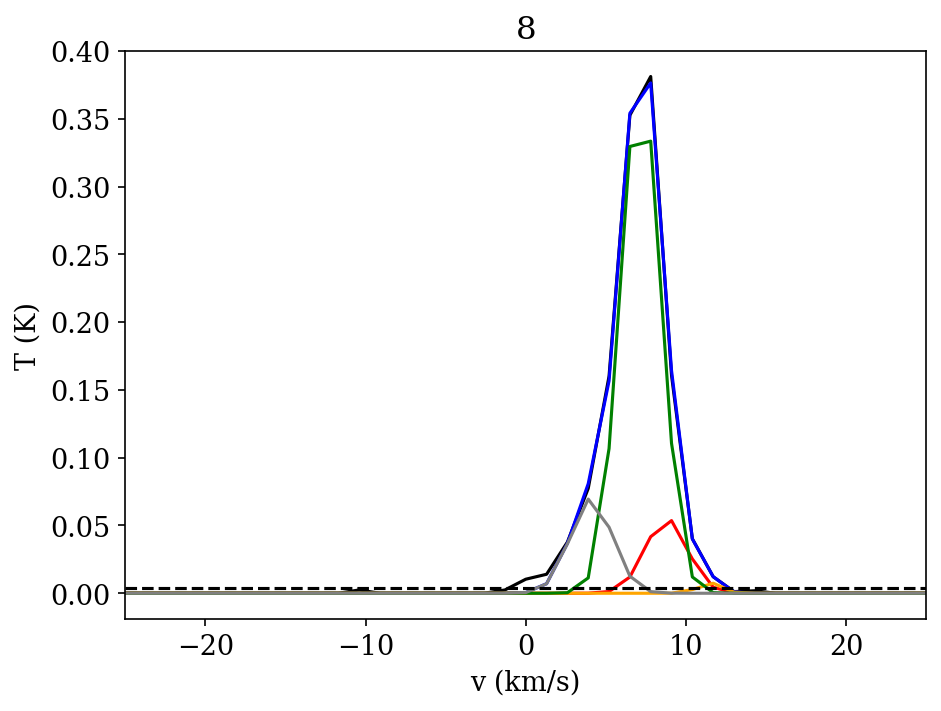}
    \includegraphics[width=1\columnwidth,height=6cm]{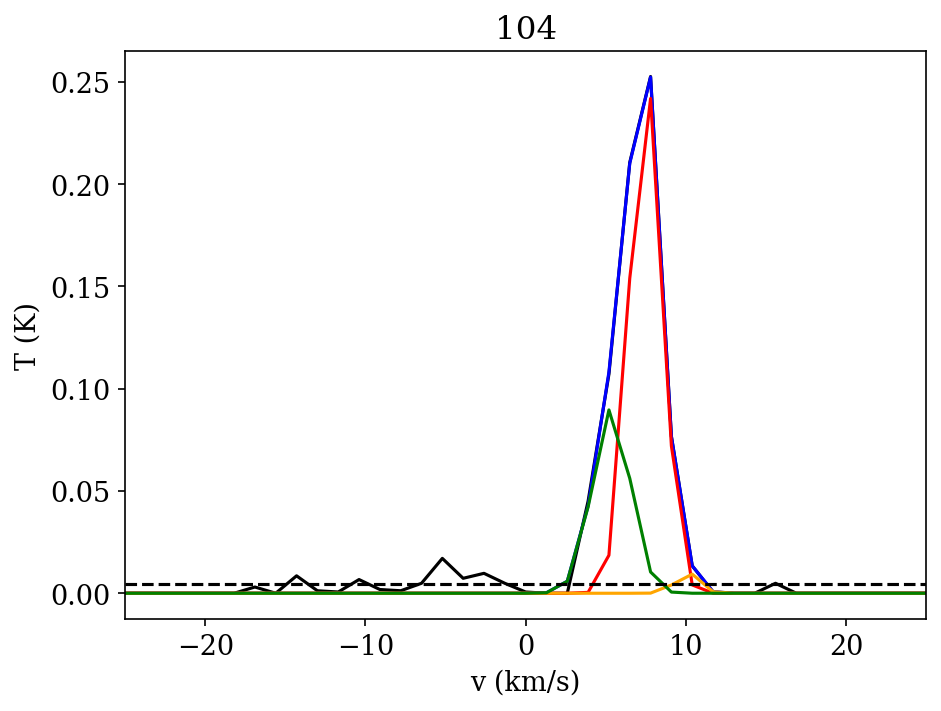}
    \includegraphics[width=1\columnwidth,height=6cm]{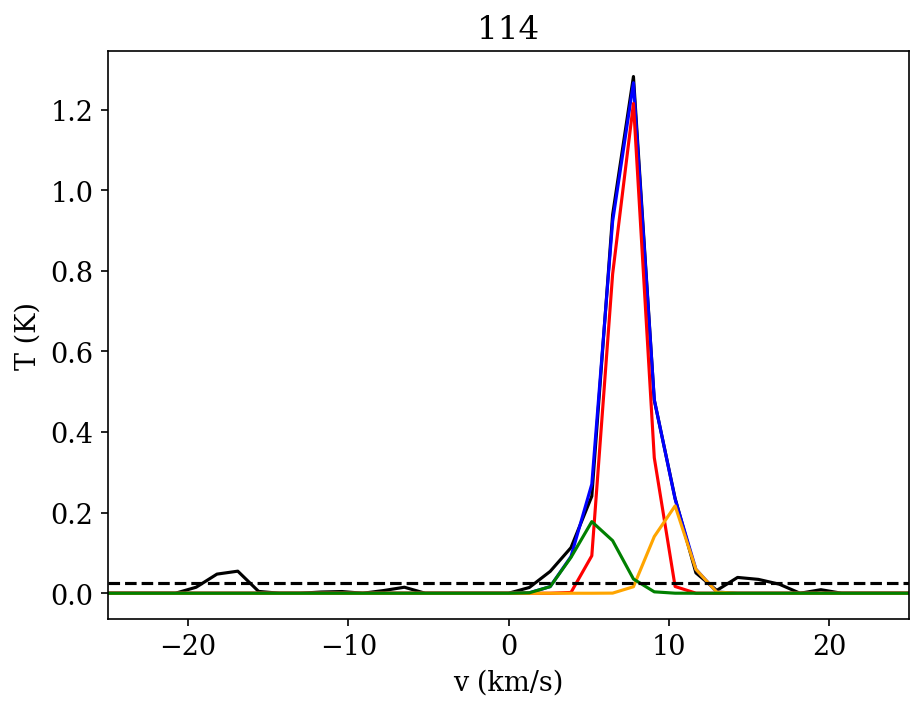}
    \caption{Median emission profiles integrated within the projected contours of three  clumps, numbered 8, 104 and 114 (black curves), and their fit by a multi-Gaussian function (blue curves). Individual Gaussian components of the function are also detailed in red, orange, green and gray. The noise level of the profile is indicated by a dashed horizontal line. The clump 114 has a smaller angular size than the two others, which explains its higher noise level.}
    \label{fig:fit_multigaussian}
\end{figure}

 We performed a quadri-Gaussian fit of the median profile after having imposed among the four components the two velocities found at the previous step using a sliding window. We allowed them to vary within a narrow, 0.5\,{\kms} wide interval. This process allowed us to identify between one and two additional potential velocity components, as illustrated in Fig.~\ref{fig:fit_multigaussian} for three different clouds, chosen as examples of the various configurations. More precisely, we first allowed two additional components, and, if one of them was found to be redundant with another component, we removed it and allowed for only one additional Gaussian. These discrete fitted components (between 1 and 4 values) were used in the same way as the sliding windows for the computation of the {\it presence} ratio,  that is for the same sky area and criteria of selection.  The mean velocities of the third and fourth Gaussian components, generally much weaker than the two first, were allowed to vary in a larger interval than the first two ($\pm 30$\,{\kms}) as they were the additional components that do not have any strong priors. We also limited the standard deviations of the Gaussians to a maximum of 1.3\,{\kms}. \\ 

 \paragraph{Final velocity selection.}

 We selected among the up to four fitted velocities the component with the highest product of the two {\it presence} ratios $PP_i = P_{ECi} \times P_{CCi}$, which we call {\it presence} product. The use of the product allowed us to give more weight to the CC over the EC by comparison with the simple use of the EC's {\it presence} ratio. We note that this quantity is no longer representing a kind of probability and is smaller than the two {\it presence} ratios.  We rejected the assignment in case the highest {\it presence} ratio for the CC (resp. EC) was found to be lower than the ratio between the CC (resp. EC) projected area and the total area WA, which is the expected ratio for a random presence distribution, as in the case of noise or unrelated very wide foreground cloud covering the whole WA and without any actual link with the EC. As we set the total WA area as four times the CC area, the rejection threshold for $P_{CCi}$ is always 0.25, while it varies from clump to clump for $P_{ECi}$. We also retained components that were located up to 2.6\,{\kms} away from the selected component and having a {\it presence} product higher than 75\% of the selected component's {\it presence} product, if they were found to exist. This allows to take into account velocity gradients in neighboring clouds. In that case, we finally assigned to the clump a velocity equal to the mean of the retained components weighted by their {\it presence} criteria.

\begin{figure}
    \centering
    \includegraphics[width=0.49\columnwidth]{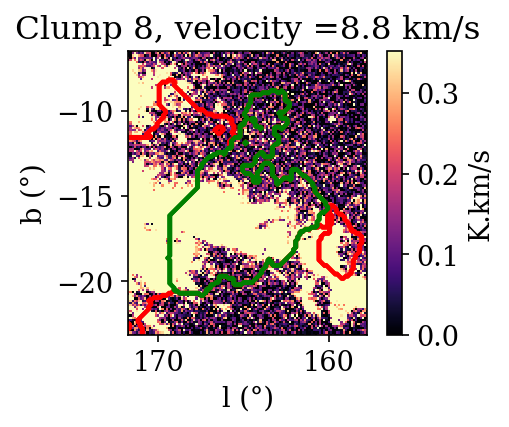}
    \includegraphics[width=0.49\columnwidth]{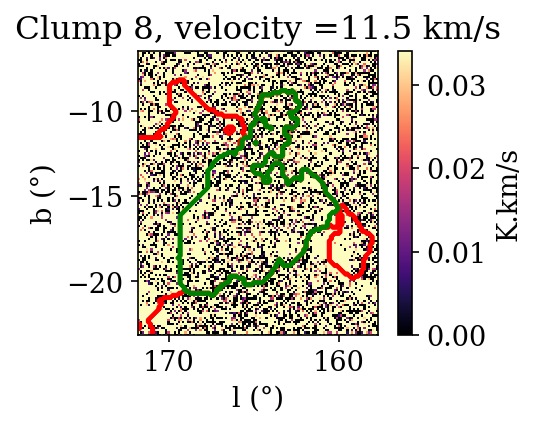}
    \includegraphics[width=0.49\columnwidth]{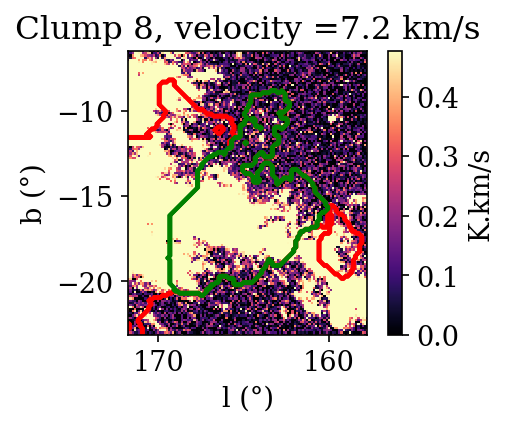}
    \includegraphics[width=0.49\columnwidth]{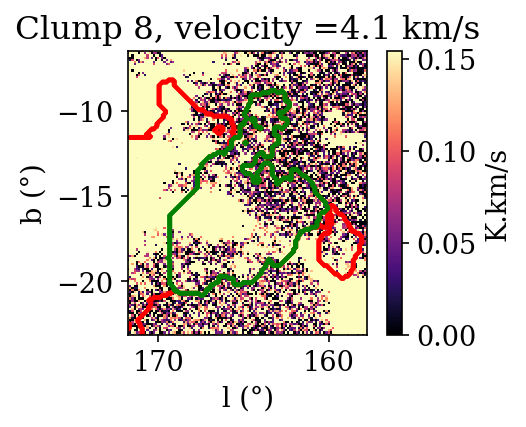}
    \caption{Integrated emission between the velocity bounds of each Gaussian component (central velocity $\pm$ the standard deviation) for the clump 8 (central velocities 8.8, 11.5, 7.2 and 4.1\,{\kms} respectively). The yellow upper limit of the color scale corresponds to the emission level at which the component is considered to be present in the line of sight. The CC clump contours are shown in green and the contours of the neighboring clouds constituting EC are shown in red.
    }
    \label{fig:presence_A08}
\end{figure}

\begin{figure}
    \centering
    \includegraphics[width=0.49\columnwidth]{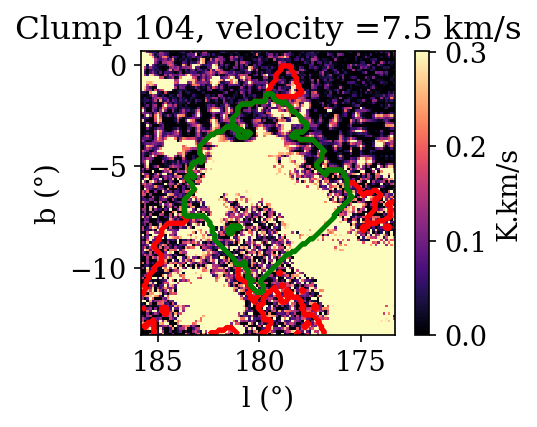}
    \includegraphics[width=0.49\columnwidth]{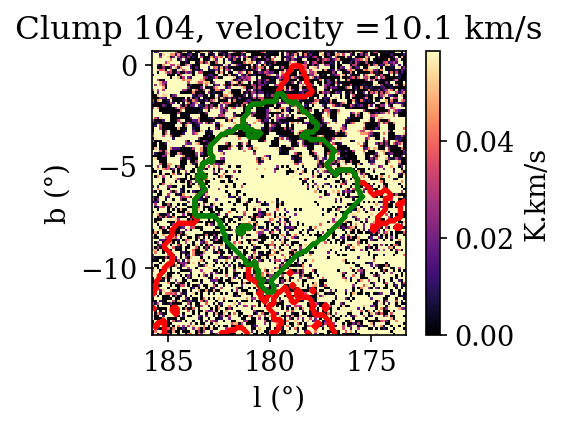}
    \includegraphics[width=0.49\columnwidth]{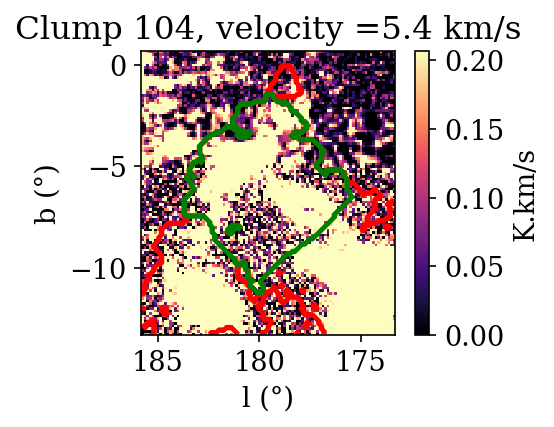}
    \caption{Same as Fig.~\ref{fig:presence_A08} for the clump 104 (velocities 7.5, 10.1 and 5.4\,{\kms}). 
    }
    \label{fig:presence_A0104}
\end{figure}

\begin{figure}
    \centering
    \includegraphics[width=0.49\columnwidth]{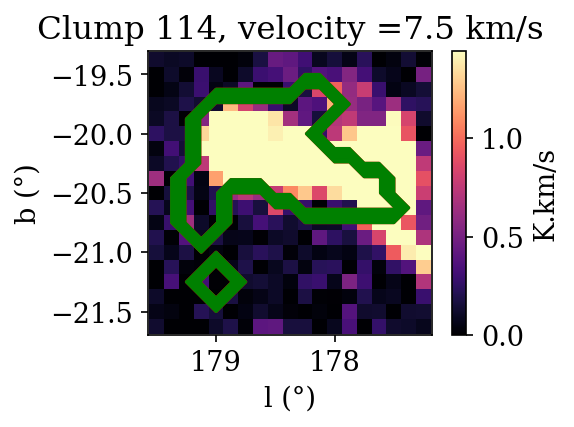}
    \includegraphics[width=0.49\columnwidth]{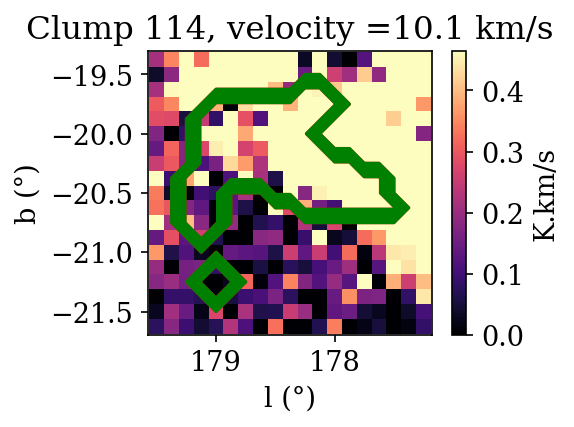}
    \includegraphics[width=0.49\columnwidth]{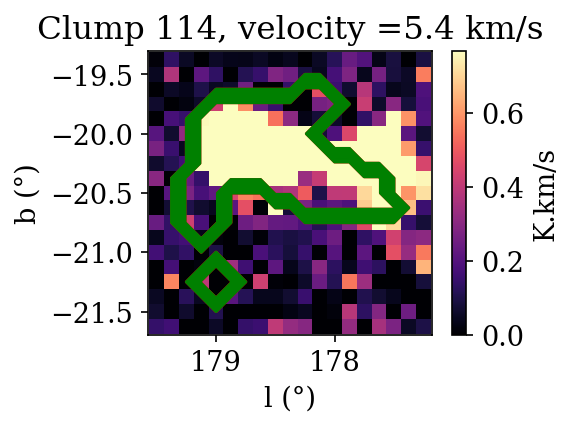}
    \caption{Same as Fig.~\ref{fig:presence_A08} for the clump 114 (velocities 7.5, 10.1 and 5.4\,{\kms}). 
    }
    \label{fig:presence_A0114}
\end{figure}

Fig.~\ref{fig:presence_A08}, ~\ref{fig:presence_A0104} and ~\ref{fig:presence_A0114} illustrate  the evaluation of the presence or absence in each line of sight of the fitted velocity intervals for the same three clouds whose fitted profiles are displayed in Fig.~\ref{fig:fit_multigaussian}. As said above, the final {\it presence} product uses the two separate presence ratios, one taking the close neighborhood (EC) into account, and one taking the considered clump (CC) only. In Fig.~\ref{fig:presence_A08} (clump 8), we can observe that the very weak 11.5\,{\kms} component can be discarded since there is no resemblance at all with the CC projection and its {\it presence} product is only of 0.12. In the case of the 4.1\,{\kms} component, its emission region covers only a small fraction of the CC projection and is marginally present outside CC or EC (grainy aspect), suggesting that an emission at a slightly shifted velocity must be present and is responsible for the low {\it presence} product of 0.13. A contrario, the 7.2 and 8.8\,{\kms} components, with their criteria of respectively 0.24 and 0.26, despite their far from perfect correspondences with the EC area, appear to cover most of it. The 8.8\,{\kms} is the component with the best CC and EC presence ratios, but both components will be used to assign its velocity to the clump 8. In Fig.~\ref{fig:presence_A0104} (clump 104), the 10.5\,{\kms} component can be discarded since there is no resemblance at all with the CC nor EC projections. The component at 7.5\,{\kms} is the one with the best {\it presence} product (0.34). Here the selection is mainly influenced by the  projection EC group of neighboring clouds, while the CC clump projection itself is not tightly fitted.  Here we assigned the velocity 7.5\,{\kms} alone because the 5.4\,{\kms} velocity's {\it presence} product is slightly beneath the threshold to be also retained (0.25 against a threshold of 0.26). In the opposite way, in Fig ~\ref{fig:presence_A0114} (clump 114), the component at 5.4\,{\kms} is the one with the best presence {\it presence} product (0.52, above the 7.5\,{\kms}  {\it presence} product of 0.44). There are no neighbors there, which allows our method to reject with confidence the component at 10.1\,{\kms}. In summary, for these three different clumps, the CC (resp. EC) {\it presence} ratios are  0.37 (resp. 0.70) for the clump 8, 0.38 (resp. 0.90) for the clump 104, and 0.72 (resp. 0.72) for the clump 114.

These examples and discussions illustrate how the {\it presence} product depends on the configurations and on the presence or absence of neighboring clumps. Isolated CCs tend to have greater criteria, while extended ECs lower the {\it presence} product. It is not possible to consider the {\it presence} product as an unified measurement available for an inter-comparison of assignment among the various clouds. Nonetheless, in general the greater the {\it presence} product, the greater our confidence in the assignation, and a high {\it presence} product ensures that both the CC and EC {\it presence} ratios are significant. We have used the arbitrary value of 0.375 for the product and highlighted the clumps with a {\it presence} product greater than this threshold  in the figures showing the velocity distribution.

\section{Results}\label{results}


Based on the above set of criteria, 42 clouds from a total of 45 were assigned a velocity, and 30 of them have a presence ratio of 0.375 or greater.  We chose to display the results in vertical planes containing the Sun and oriented along varying Galactic longitudes, as in \cite{Ivanova21}. Those planes are shown in  Figs.~\ref{fig:coupe1}, \ref{fig:coupe2}, \ref{fig:coupe3} and \ref{fig:coupe4}. Based on the characteristic velocity assigned to each cloud, we computed a distance and a velocity for a series of directions within the cloud contours taken from a longitude-latitude grid eight times looser than the CO grid (grid steps of 0.125\fdeg~ in longitude and 1.0\fdeg~ in latitude). This grid is chosen to display a maximum of information on the structures in the map while keeping the figure readable. 
The distance is defined as the barycenter of the extinction due to the cloud in the considered direction, and the velocity is derived from fitting the CO profile within the window determined during the selection step, allowing for a small velocity shift by one CO channel. This allows to make visible weak velocity gradients within the cloud. These points are marked by thick crosses with black contours on Figs.~\ref{fig:coupe1}, \ref{fig:coupe2}, \ref{fig:coupe3} and \ref{fig:coupe4}.

These results for individual clumps are assembled in Table \ref{tab:results}. The table includes the three lacks of assignments. Their three direction-distance pairs are close to those of three neighboring clouds with assigned velocities, suggesting that they may be extensions of these clouds. We have seen in section \ref{dustmaps} that the Fellwalker decomposition may separate adjacent structures sharing the same velocity. In this case, the presence ratio criterion may fail. This could also correspond to structures or artifacts in the inverted extinction map ({\it Fingers of god} effects) or clouds devoid of CO and not eliminated by the mask. The results for the full grid (thick crosses in the figures) are listed  in an additional table available from the CDS whose first lines are shown in Table~\ref{tab:points}.

\begin{figure}
    \centering
    \includegraphics[width=1\columnwidth, height=2.5cm]{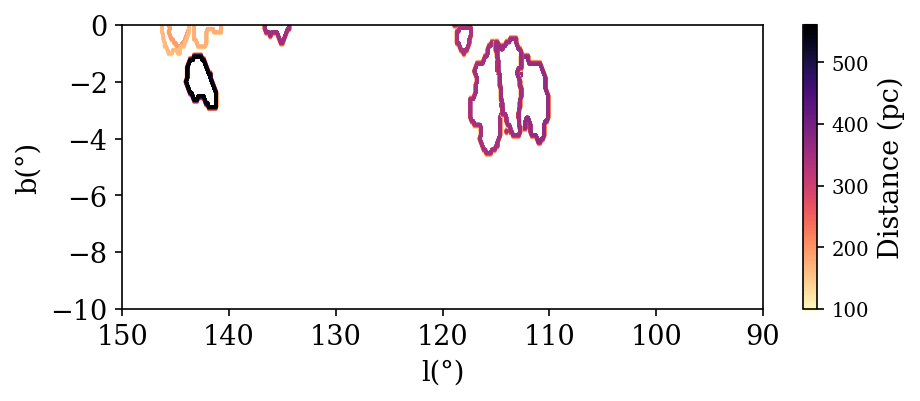}
    \caption{Contours of the clump in our map outside of the main Taurus region. The map is in Galactic coordinates.
    }
    \label{fig:contours_clumps_etendu}
\end{figure}

\begin{figure*}
    \centering
    \includegraphics[width=\textwidth]{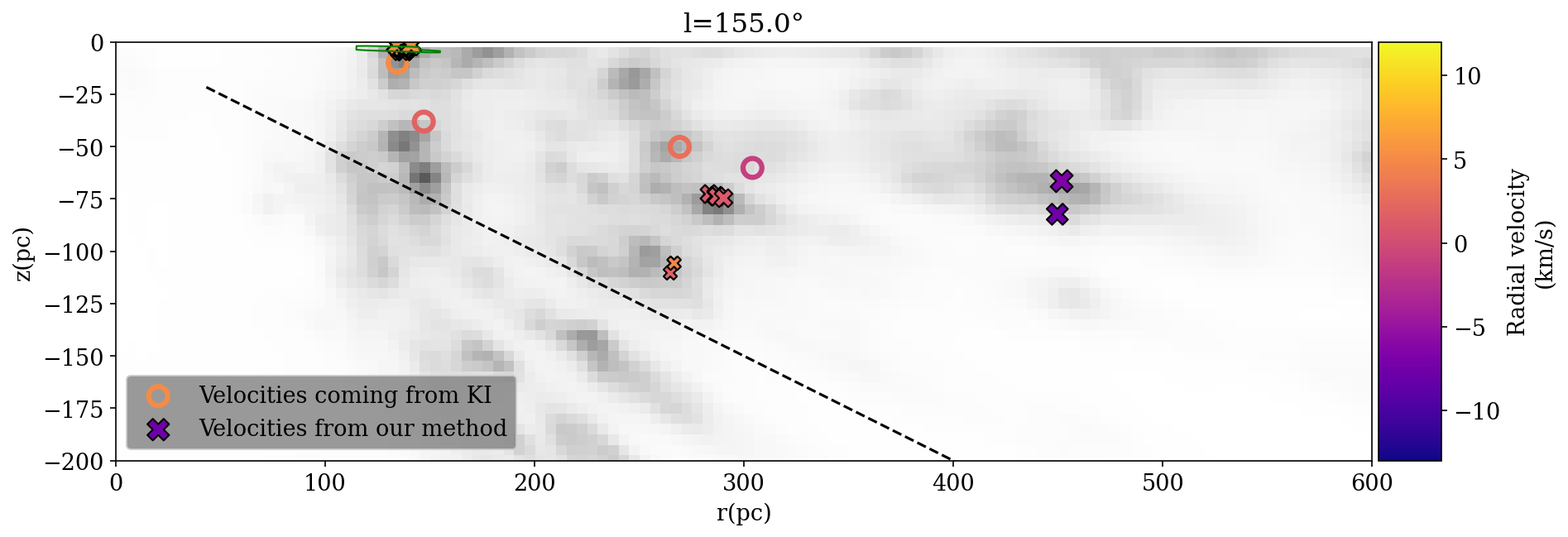}
    \includegraphics[width=\textwidth]{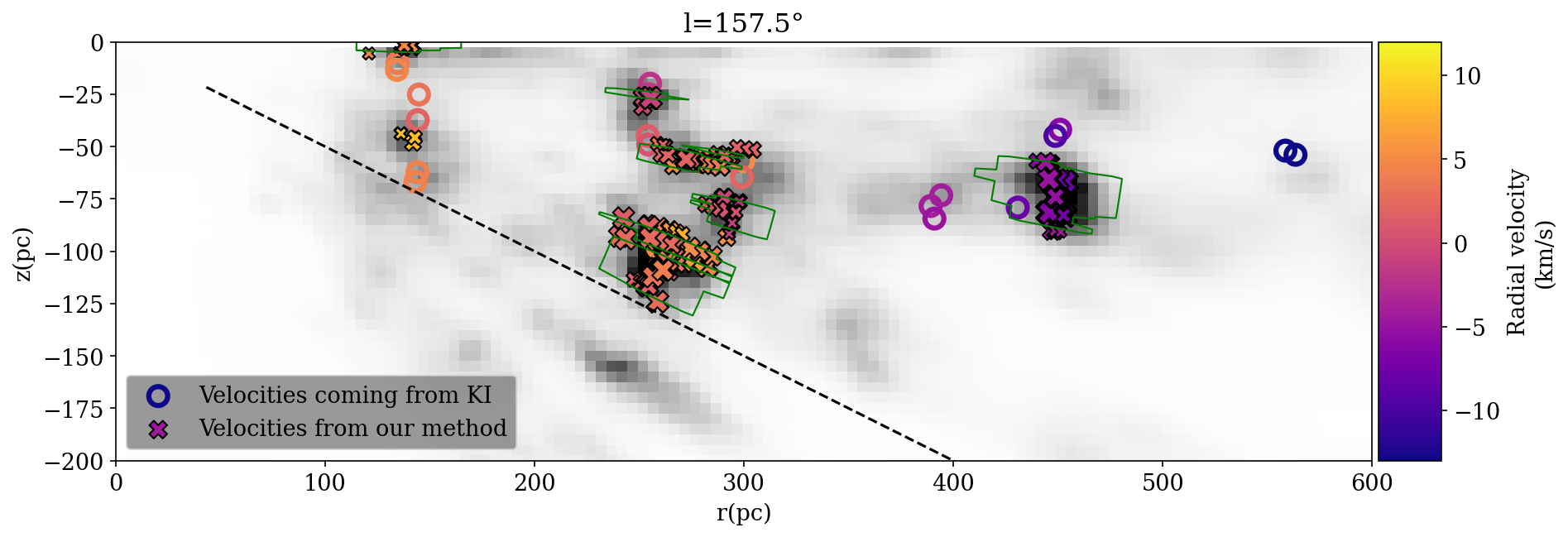}
    \includegraphics[width=\textwidth]{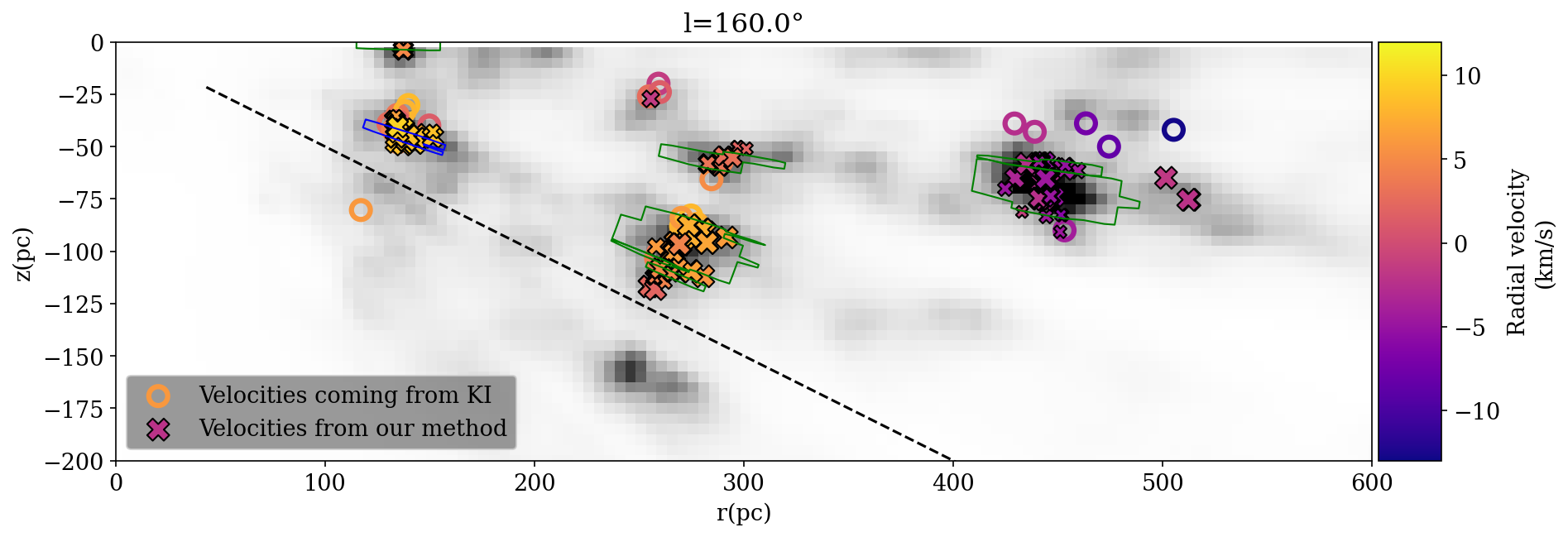}
    \caption{Comparison between radial velocities assigned to clouds with our method (marked with crosses) and those derived from KI spectra and attributed by \cite{Ivanova21} to dust structures located in the same region of 3D space (marked with circles). We show the results in vertical planes containing the Sun and oriented towards the longitude l indicated at top of the figures. As in paper I,  all data points in the interval $[l-1.25^\circ,l+1.25^\circ]$ are treated as if they were contained in the l plane. In shades of grey is the extinction density map at the given longitude. Our points are positioned at the barycenter of the clumps in the considered lines of sight, and the mean velocities are reported via the color bar. Points based on interstellar KI and initially  {\it manually} located by \cite{Ivanova21} close to the assigned clouds are kept at the same positions. In blue are the contours of the Fellwalker clumps in the given slice that have an assigned velocity and a presence product lower than 0.375, and in green are the ones with presence product greater than 0.375, while non-assigned clump are in black. The dotted line marks the $b=-28^\circ$ limit of the CO emission spectral cube. In a given clump, the sampling is every $1.0^\circ$ in b and $0.125^\circ$ in l. Empty crosses are used where the Fellwalker clump is only 1 voxel deep, preventing to calculate any barycenter.}
    \label{fig:coupe1}
\end{figure*}

\begin{figure*}
    \centering
    \includegraphics[width=\textwidth]{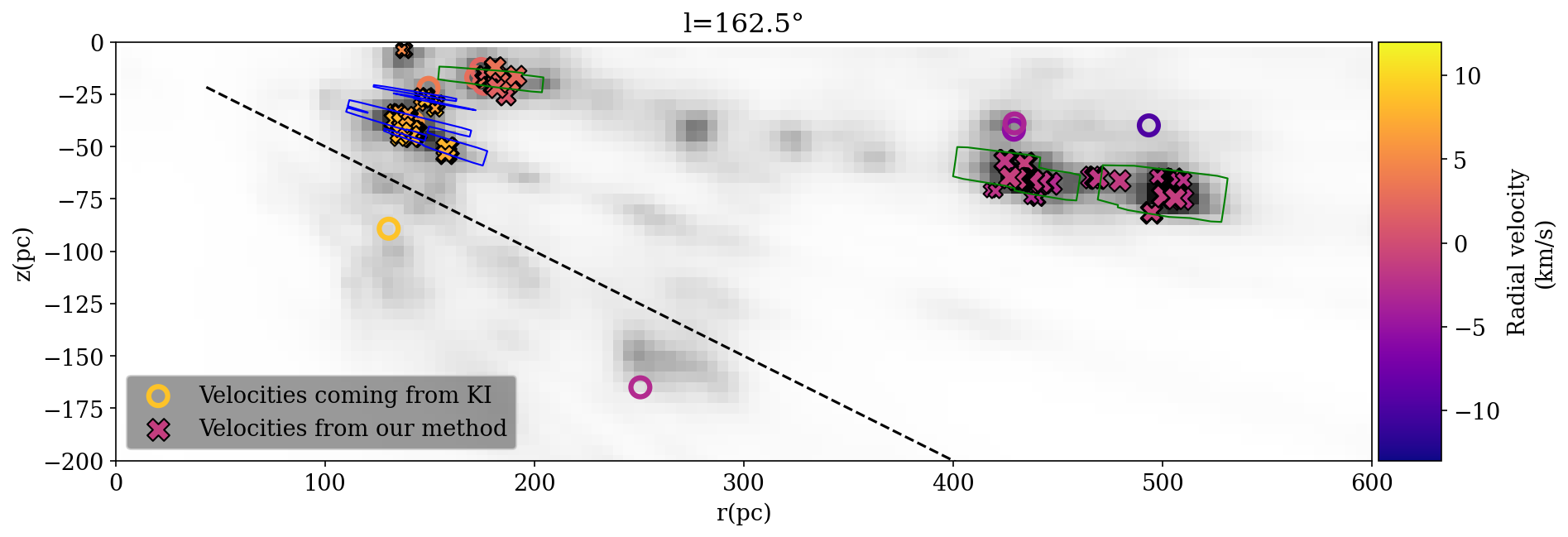}
    \includegraphics[width=\textwidth]{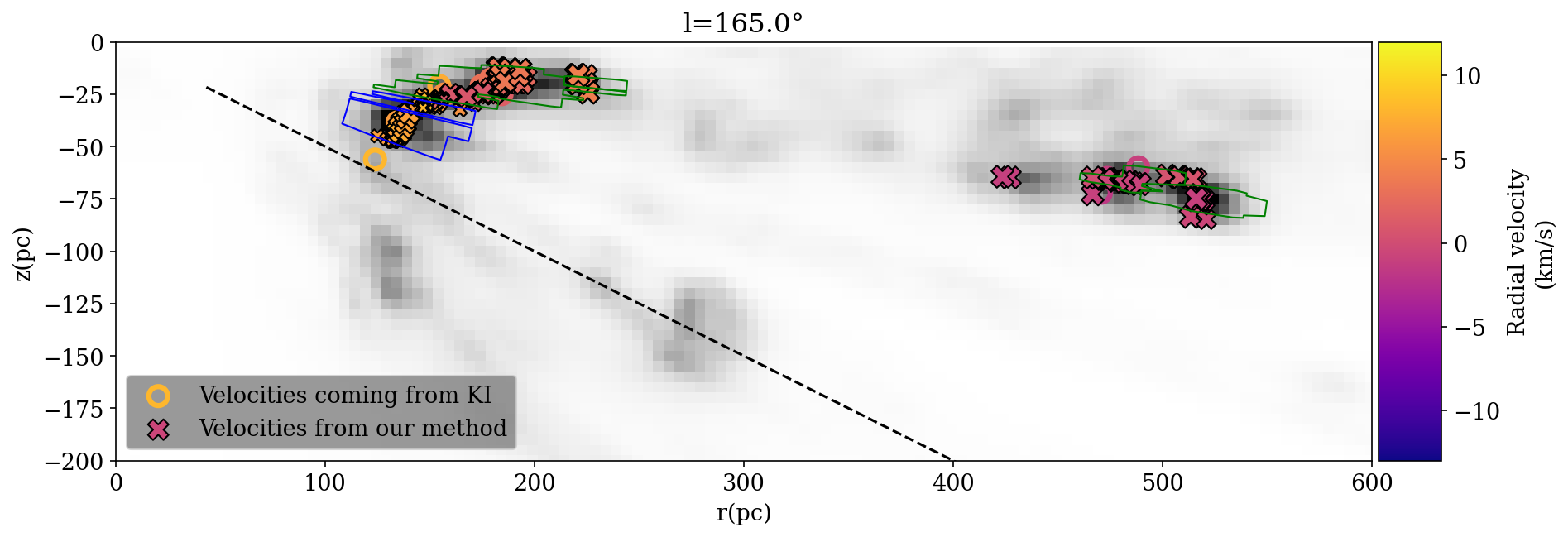}
    \includegraphics[width=\textwidth]{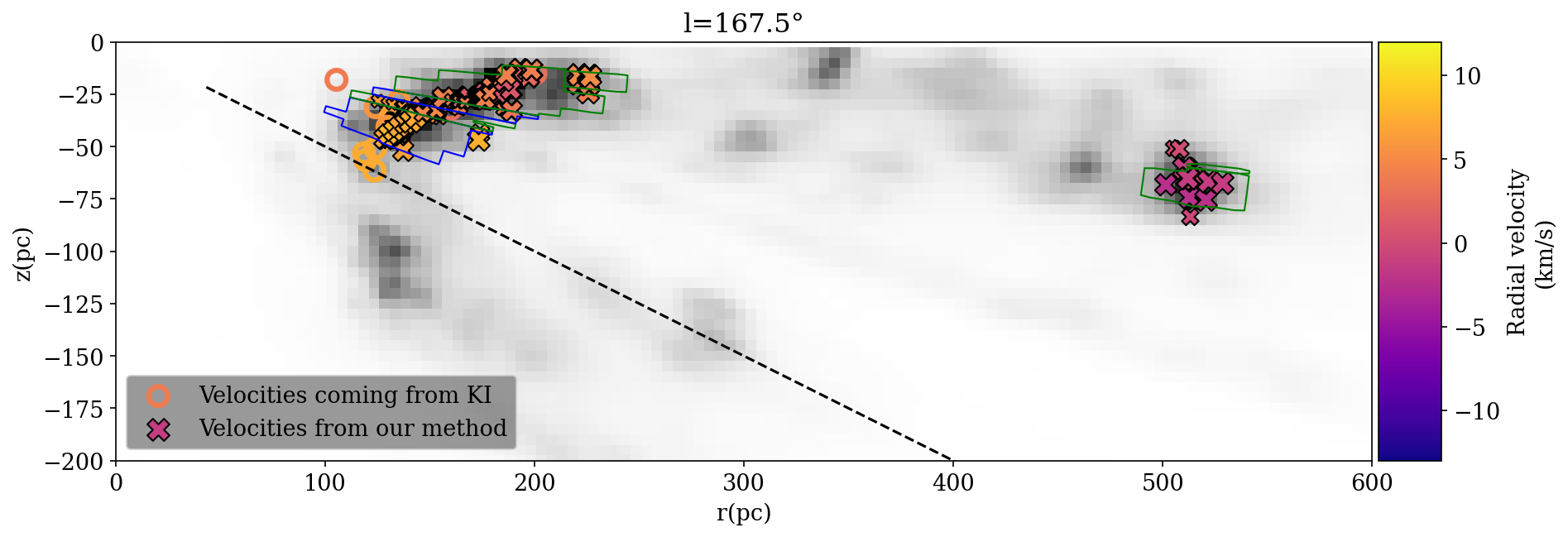}
    \caption{Same as Fig.\ref{fig:coupe1} for longitudes between $161.25^\circ$ and $168.75^\circ$.}
    \label{fig:coupe2}
\end{figure*}

\begin{figure*}
    \centering
    \includegraphics[width=\textwidth]{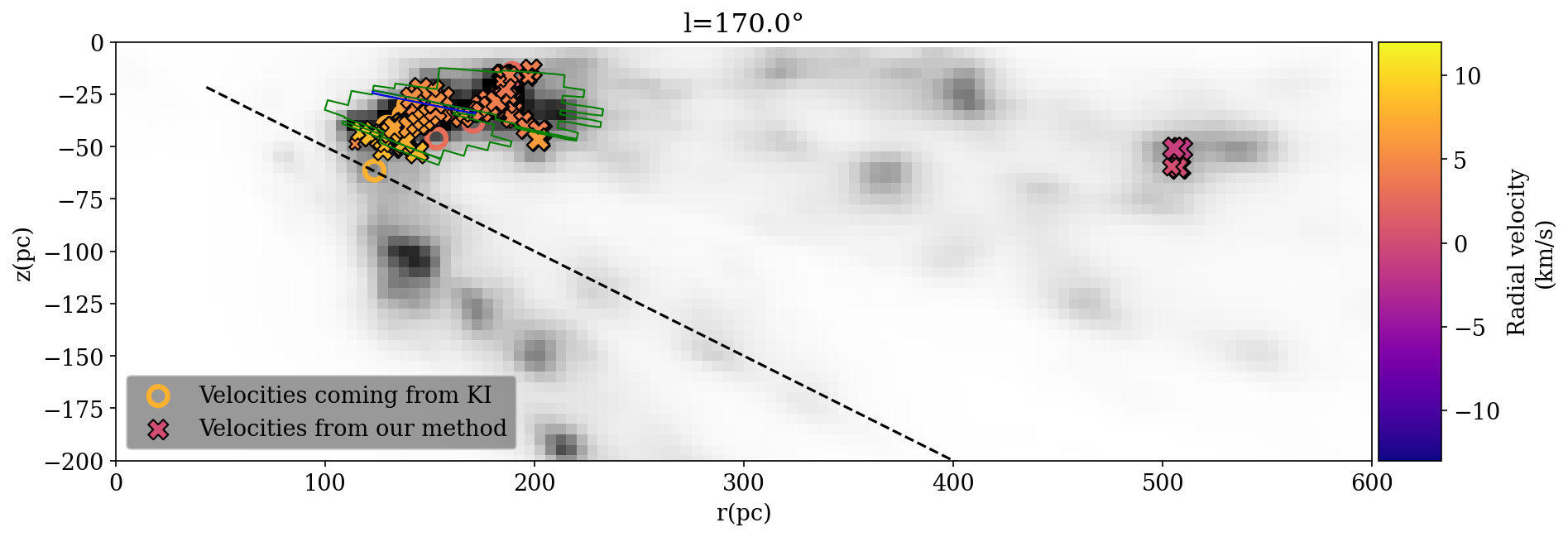}
    \includegraphics[width=\textwidth]{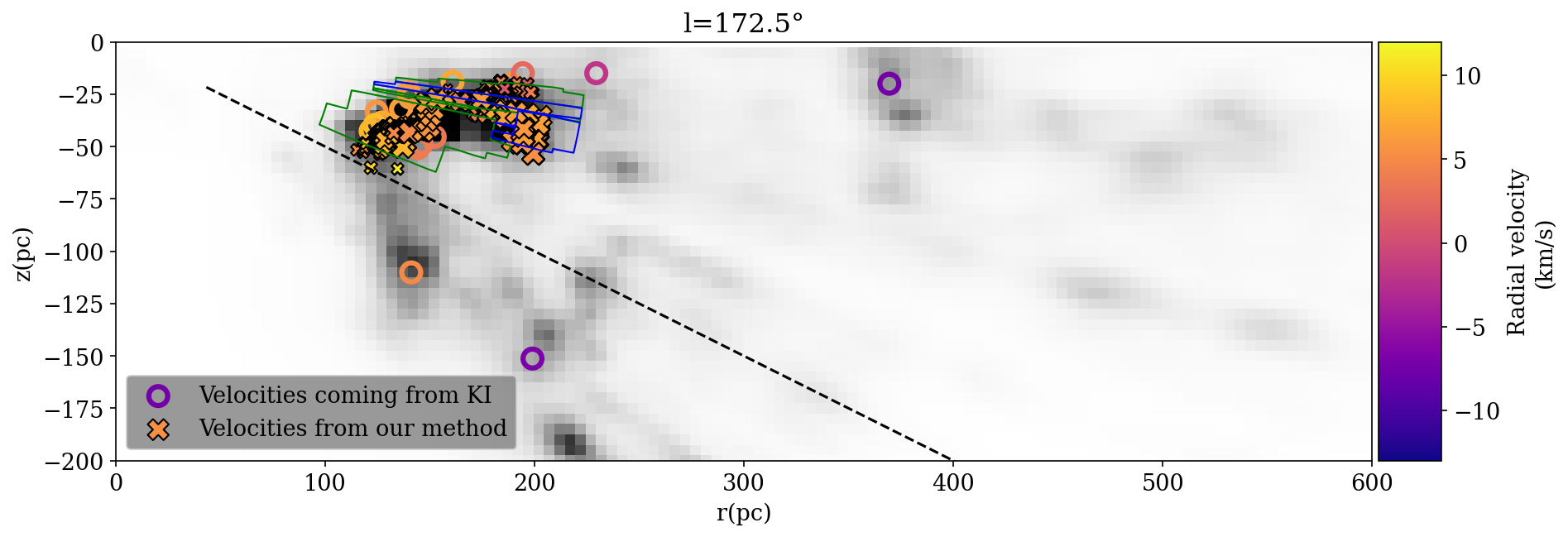}
    \includegraphics[width=\textwidth]{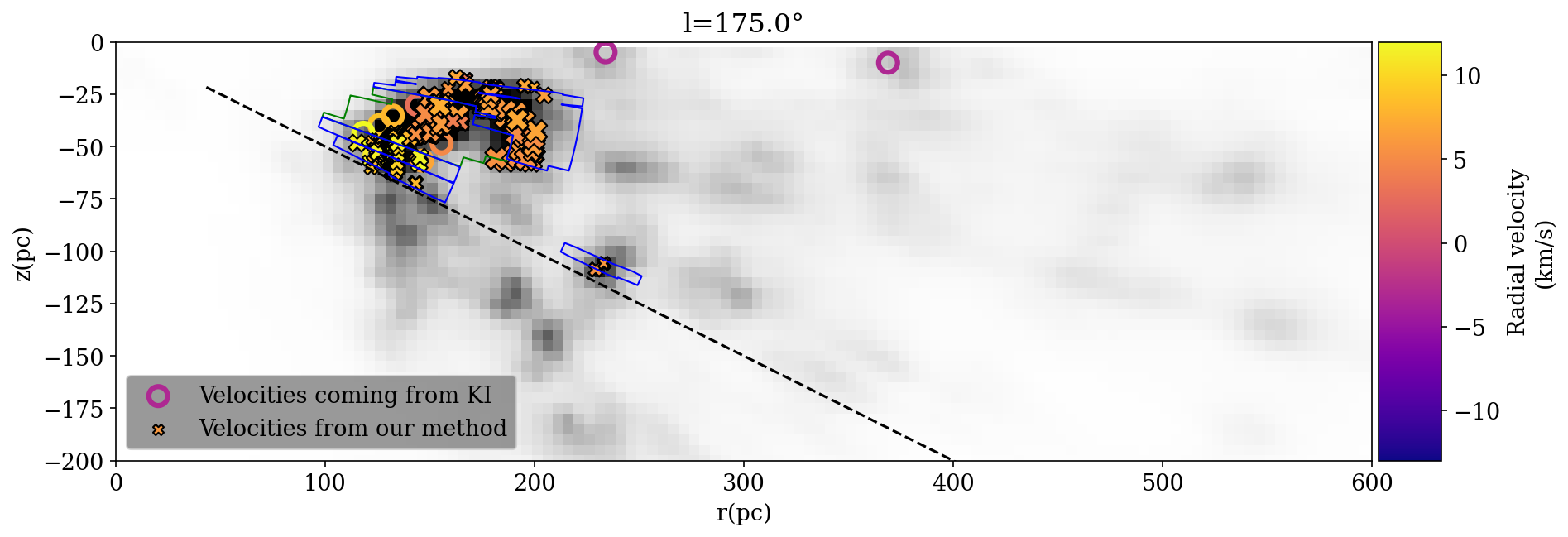}
    \caption{Same as Fig.\ref{fig:coupe1} for longitudes between $168.75^\circ$ and $176.25^\circ$.}
    \label{fig:coupe3}
\end{figure*}

\begin{figure*}
    \centering
    \includegraphics[width=\textwidth]{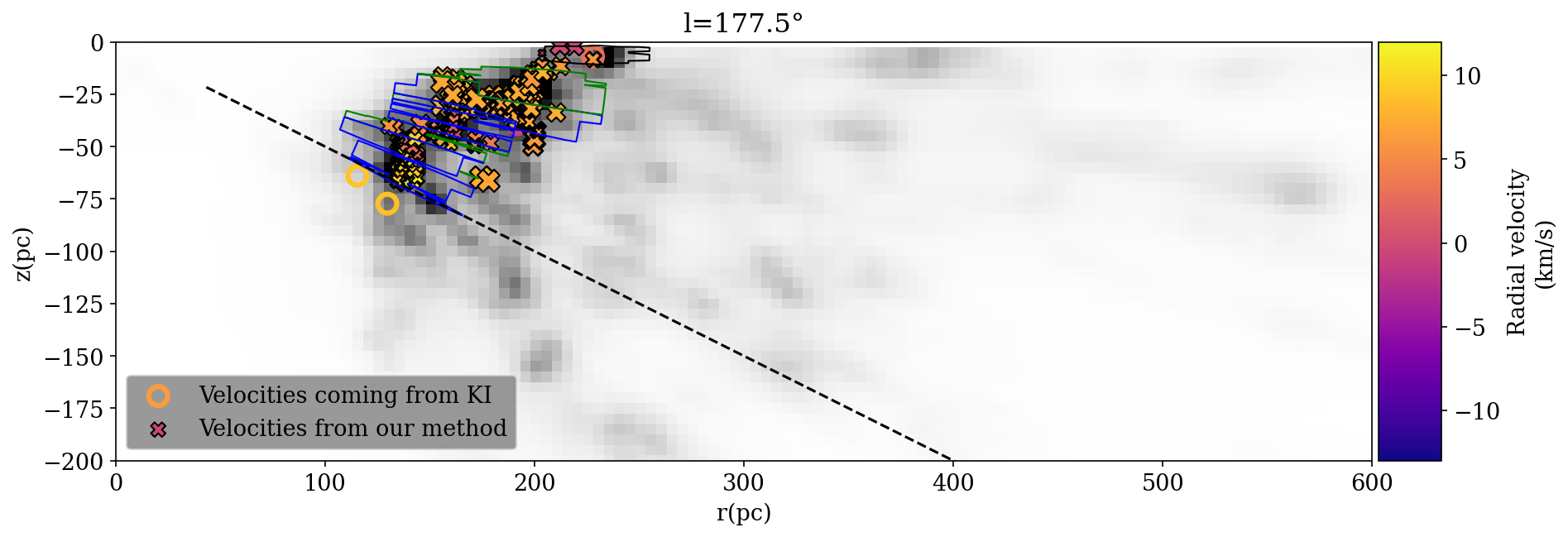}
    \includegraphics[width=\textwidth]{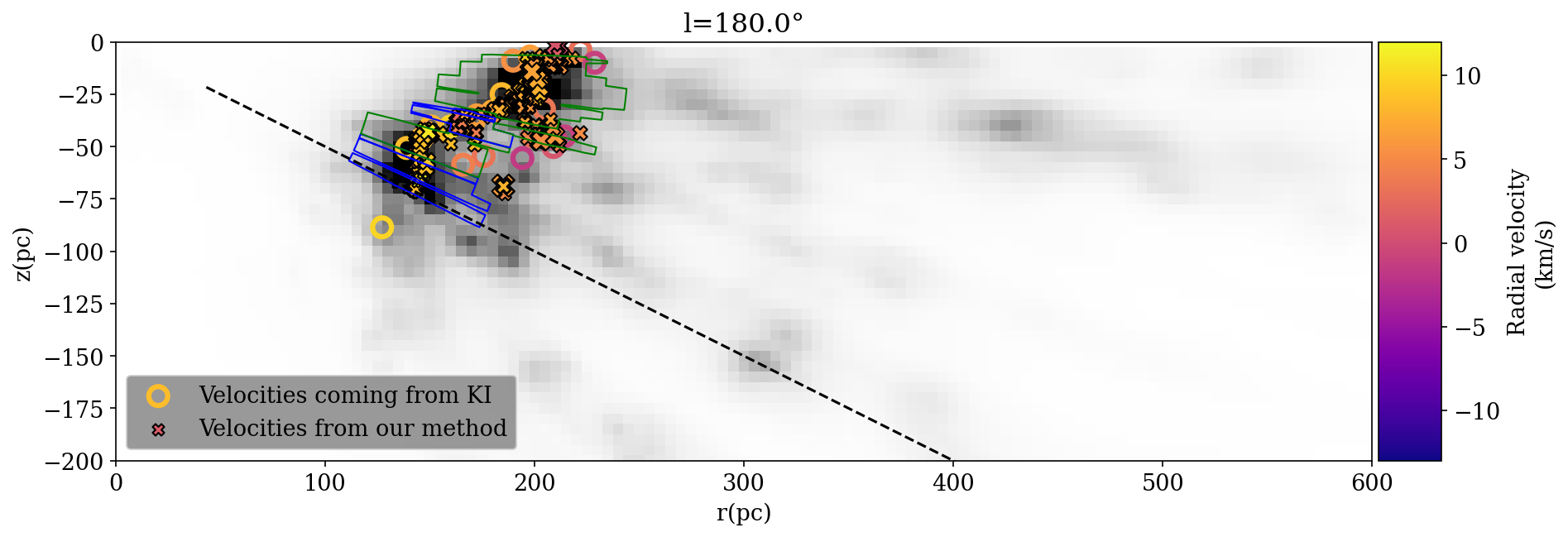}
    \includegraphics[width=\textwidth]{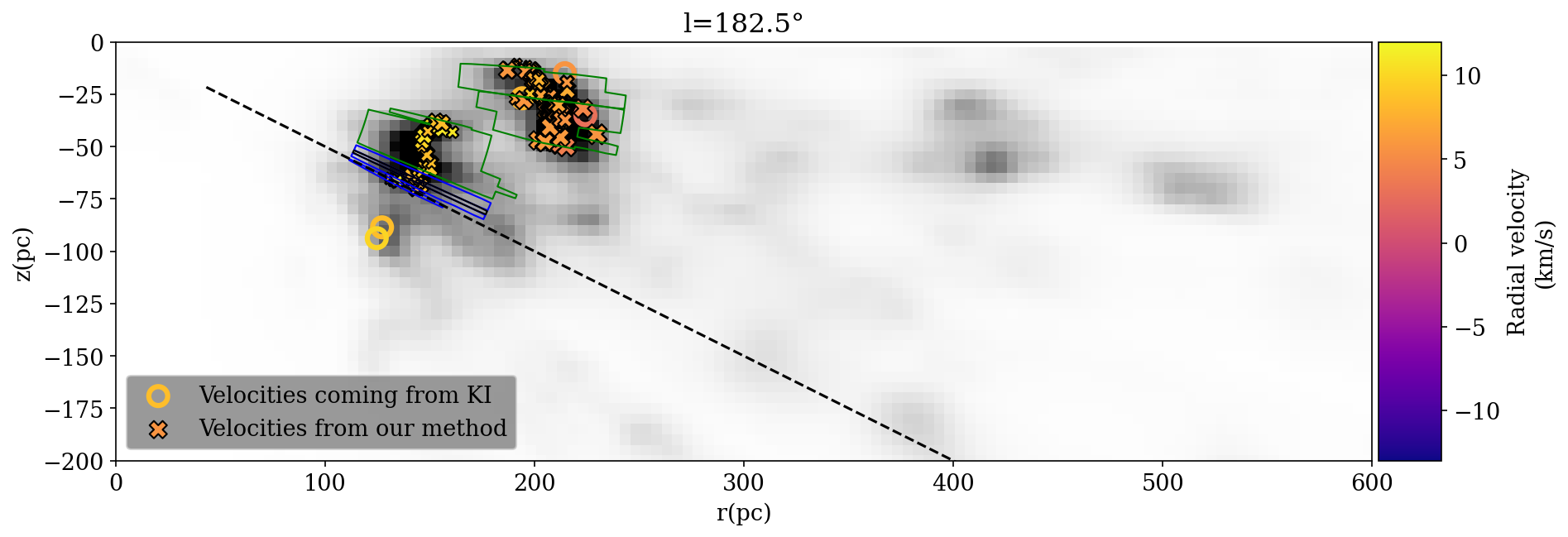}
    \caption{Same as Fig.\ref{fig:coupe1} for longitudes between $176.25^\circ$ and $183.75^\circ$.}
    \label{fig:coupe4}
\end{figure*}

\begin{figure*}
    \centering
    \includegraphics[width=\textwidth]{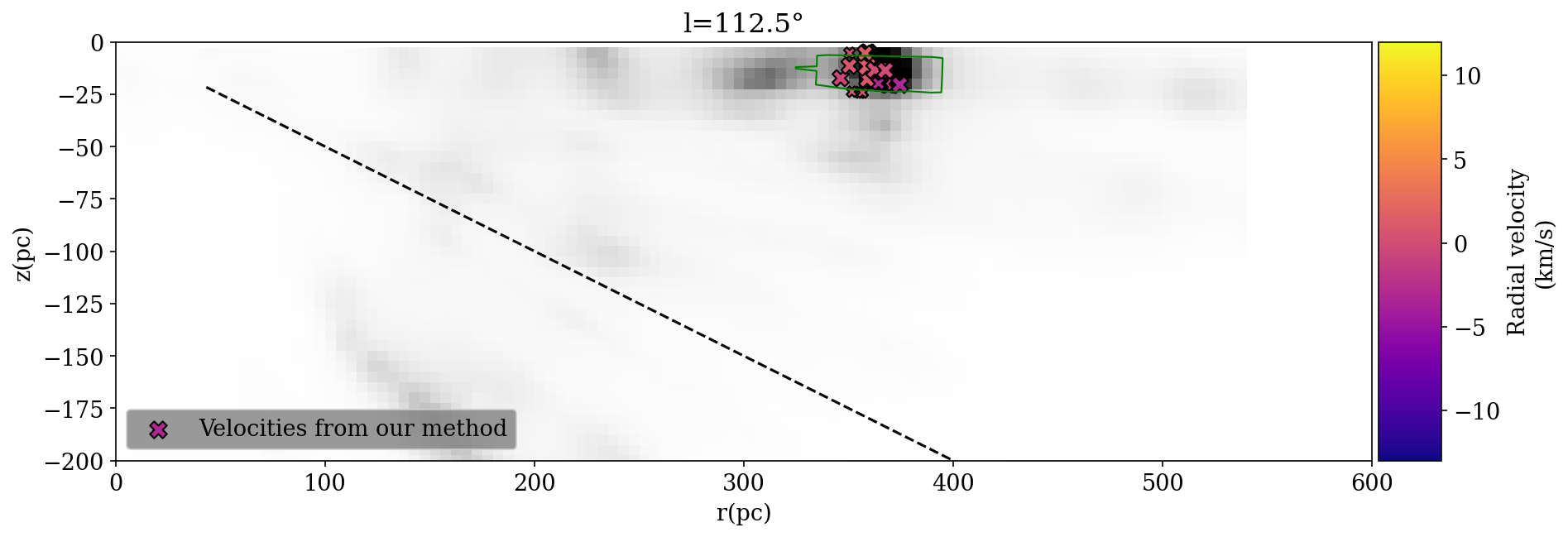}
    \includegraphics[width=\textwidth]{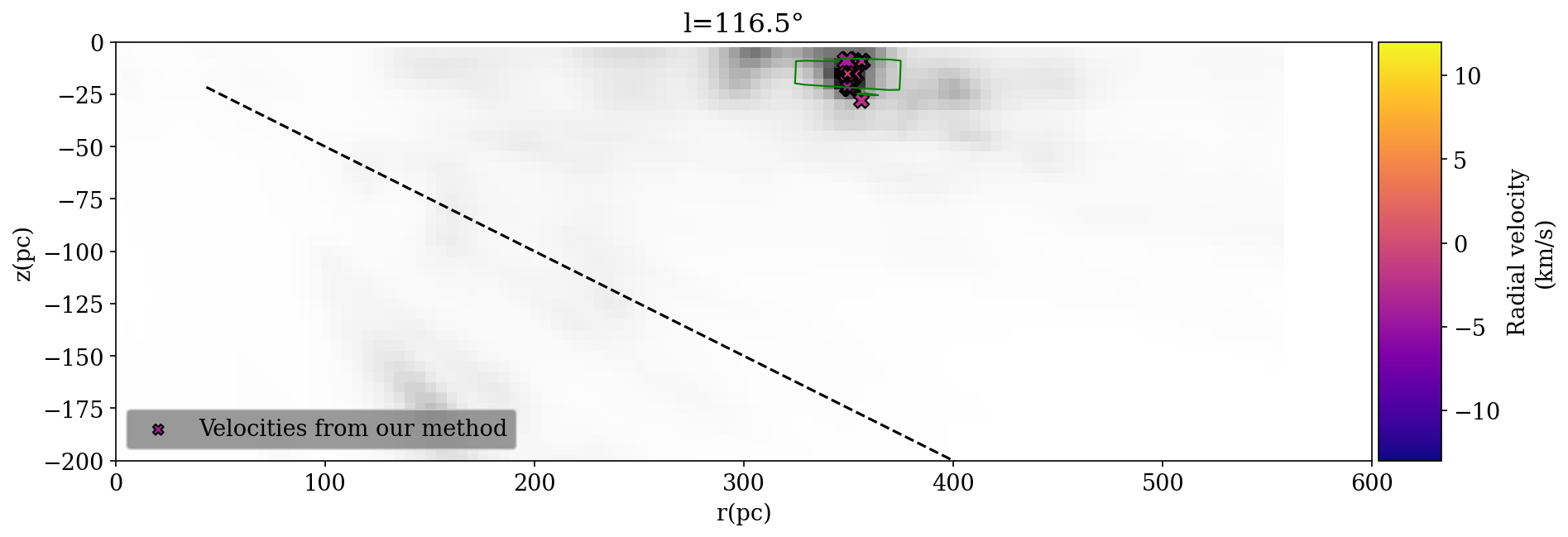}
    \includegraphics[width=\textwidth]{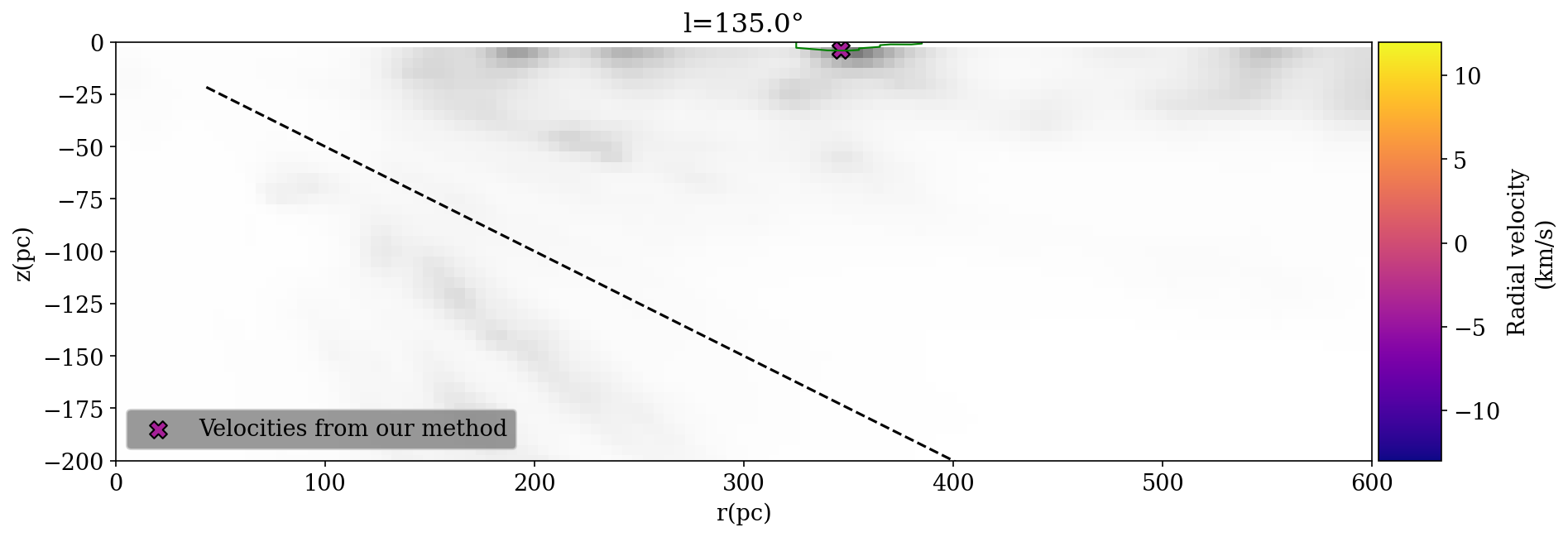}
    \caption{Similar to Fig.\ref{fig:coupe1} for longitudes less than $150^\circ$ featuring clumps with an assignated velocity.}
    \label{fig:coupeet1}
\end{figure*}

\begin{figure*}
    \centering
    \includegraphics[width=\textwidth]{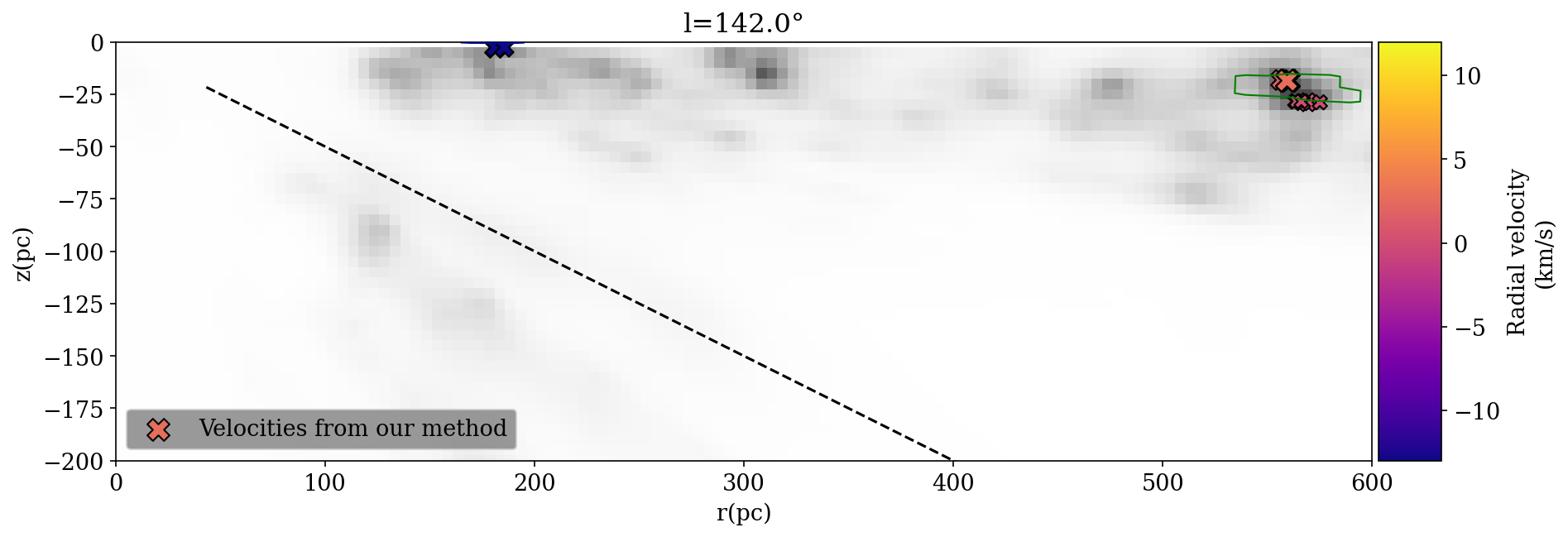}
    \includegraphics[width=\textwidth]{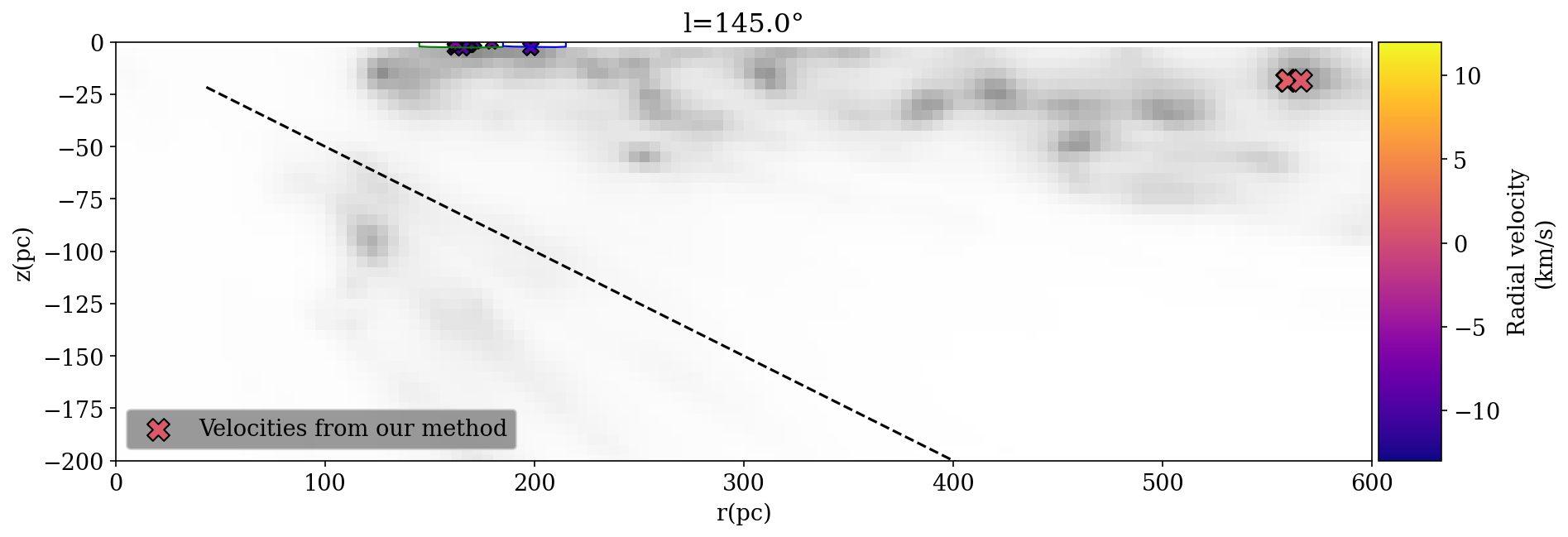}
    \caption{Similar to Fig.\ref{fig:coupe1} for longitudes less than $150^\circ$ featuring clumps with an assignated velocity.}
    \label{fig:coupeet2}
\end{figure*}

Although we focused on the Taurus area, the parallelepipedic shape of the volume we extracted from the full 3D extinction density distribution is such that the longitude range extends to 90$^\circ$. As shown in Fig.\ref{fig:contours_clumps_etendu}, that area contains way less clouds, and those are quite near the galactic plane. Since they were assigned a velocity, they are presented here in Fig.\ref{fig:coupeet1} and \ref{fig:coupeet2}.

\begin{table*}[b]
    \centering
    \begin{tabular*}{\textwidth}{@{\extracolsep{\fill}} *{11}{c}}
        clump ID & cloud & dmean &  bmean &  lmean& dsig & bsig & lsig &  <v> & vsig & $PP_i$\\
         & & (pc) & ($^\circ$) & ($^\circ$) & (pc) & ($^\circ$) & ($^\circ$) & ({\kms}) & ({\kms}) & \\
        \hline
      3 & Taurus      &       155 &     -24.35 &     178.95 &  16 &  1.33 &  2.42 &         9.8 &   1.6 &  0.311 \\
      7 & Taurus      &       138 &     -19.95 &     169.91 &  12 &  0.43 &  0.52 &         8.3 &   1.4 &  0.395\\
      8 & Taurus      &       144 &     -15.88 &     165.55 &  13 &  2.38 &  2.09 &         8.0 &   1.5 &  0.263 \\
      9 & Taurus      &       150 &     -15.31 &     172.77 &  17 &  2.87 &  2.51 &         6.5 &   1.1 &  0.428 \\
      15 & Taurus      &       145 &     -20.60 &     176.45 &  14 &  1.11 &  1.15 &        11.3 &   1.7 &  0.314 \\
      17 & Taurus      &       153 &     -20.09 &     179.22 &  15 &  1.07 &  0.59 &         2.1 &   0.5 &  0.237 \\
      18 & Taurus      &       157 &     -19.20 &     182.47 &  15 &  2.42 &  1.61 &        10.1 &   1.2 &  0.344 \\
      19 & Taurus      &       149 &     -17.82 &     159.44 &  12 &  0.93 &  0.60 &        10.2 &   1.9 &  0.148 \\
      25 & Taurus      &       138 &      -0.93 &     157.89 &   9 &  0.62 &  1.94 &         5.4 &   1.4 &  0.238 \\
      41 & Taurus     &       158 &     -24.73 &     182.24 &  16 &  0.18 &  0.34 &    \\
      47 & Taurus     &       161 &     -19.87 &     186.51 &  16 &  0.23 &  0.31 &    \\
      52 & Taurus      &       167 &     -13.56 &     177.76 &  15 &  1.07 &  1.15 &         2.4 &   1.0 &  0.198 \\
      59 & Taurus      &       160 &     -10.59 &     168.32 &  16 &  0.71 &  1.02 &         3.8 &   1.7 &  0.268 \\
      61 & Auriga      & 184 &      -7.55 &     168.34 &  20 &  1.88 &  2.73 &         2.5 &   1.3 &  0.346 \\
      66 & Taurus      &       187 &     -10.16 &     175.29 &  20 &  2.05 &  1.75 &         5.5 &   1.9 &  0.205 \\
      103 & Taurus    &       169 &     -11.35 &     179.09 &  11 &  0.25 &  0.28 &         5.7 &   0.6 &  0.203 \\
      104 & Taurus    &       200 &      -6.19 &     179.77 &  17 &  1.83 &  1.62 &         7.5 &   1.0 &  0.346 \\
      114 & Taurus    &       193 &     -20.24 &     178.57 &   9 &  0.36 &  0.44 &         6.4 &   1.7 &  0.517 \\
      131 & Taurus     &       215 &     -10.89 &     182.80 &  13 &  1.64 &  1.26 &         5.8 &   1.6 &  0.320 \\
      137 & Taurus    &       223 &     -10.39 &     184.28 &  14 &  0.70 &  0.55 &         4.9 &   1.0 &  0.325 \\
      141 & Taurus    &       210 &      -0.99 &     178.77 &  11 &  0.38 &  0.36 &        -0.2 &   1.2 &  0.355 \\
      150 & Taurus & 228 &  -1.60 &     177.40 &  11 &  0.49 &  0.30 & &\\
      149 & Auriga & 224 &      -4.65 &     166.41 &   9 &  0.72 &  0.79 &         3.6 &   1.0 &  0.239 \\
      173 &Perseus    &       288 &     -19.84 &     159.14 &  15 &  1.20 &  0.70 &         6 &   1.7 &  0.536 \\
      174 &Perseus    &       276 &     -20.08 &     157.64 &  12 &  0.52 &  0.51 &         1.0 &   1.5 &  0.485 \\
      178 &Perseus    &       282 &     -23.08 &     158.02 &  12 &  1.03 &  0.91 &         3.2 &   1.7 &  0.414 \\
      228 &California    &       445 &      -8.64 &     161.23 &  18 &  0.73 &  1.08 &        -2.8 &   1.5 &  0.506 \\
      237 &California    &       453 &      -8.93 &     158.17 &  14 &  1.03 &  0.95 &        -6.2 &   0.8 &  0.474 \\
      242 &California    &       487 &      -7.82 &     164.32 &  14 &  0.44 &  0.58 &        -1.1 &   1.2 &  0.263 \\
      247 &California    &       507 &      -8.11 &     162.38 &  14 &  0.58 &  0.70 &        -0.6 &   1.6 &  0.341 \\
      252 &California    &       512 &      -6.50 &     168.63 &  12 &  0.51 &  0.51 &        -1.5 &   1.4 &  0.572 \\
      258 &California    &       521 &      -8.02 &     166.08 &  12 &  0.55 &  1.10 &        -0.9 &   2.0 &  0.691 \\
 \hline
 \hline
      164 & Others in Fig.2 & 256 &     -24.65 &     175.17 &   8 &  0.39 &  0.26 &         5.5 &   0.6 &  0.108 \\
      167 & " & 255 &      -5.96 &     158.18 &   8 &  0.45 &  0.47 &        -1.3 &   1.0 &  0.332 \\
      180 & "& 292 &     -11.06 &     158.55 &  16 &  0.72 &  0.87 &         3.6 &   1.3 &  0.157 \\
      183 & " & 302 &     -15.26 &     157.14 &  11 &  0.82 &  0.77 &        -0.5 &   0.4 &  0.234 \\
\hline
\hline
      199 & Others at l$\leq$150 & 344 &      -0.45 &     117.98 &  12 &  0.23 &  0.33 &         2.4 &   1.3 &  0.289 \\
      202 & "    &       359 &      -1.98 &     113.61 &  15 &  0.80 &  0.54 &         0.8 &   0.5 &  0.473 \\
      203 & "    &       352 &      -2.58 &     115.66 &  12 &  0.89 &  0.68 &        -2.4 &   1.9 &  0.357 \\
      205 & "    &       368 &      -2.49 &     111.72 &  12 &  0.70 &  0.70 &        -1.6 &   1.2 &  0.418 \\
      211 & "    &       355 &      -0.15 &     135.44 &  13 &  0.15 &  0.60 &        -2.3 &   1.1 &  0.766 \\         
      96 & "     &       168 &      -0.37 &     145.06 &  12 &  0.28 &  0.66 &        -8.1 &   1.2 &  0.232\\
      97 & "     &       181 &      -0.25 &     142.30 &  10 &  0.23 &  0.71 &       -13 &   1.7 &  0.120 \\
      262 & " & 561 &      -2.03 &     142.56 &  12 &  0.44 &  0.70 &         1.6 &   1.3 &  0.303 \\
      118 & " & 197 &      -0.29 &     144.70 &   8 &  0.22 &  0.49 &        -8.5 &   1.5 &  0.138 \\

\end{tabular*}
    
    \caption{Table of the mean coordinates and assigned velocity. The inside over total presence ratio is given as an indicator of the assignation's quality for each clump. The three clouds without assigned velocity are listed close to neighboring clouds likely responsible for the fact that criteria were not met (see text).
    \label{tab:results}
    }
\end{table*}

\section{Comparisons with previous results}\label{validation}

\subsection{Comparisons with neutral potassium absorption data}

In paper I we described a non-automated method of velocity assignments to the dust concentrations appearing in the 3D extinction density maps of the same volume \citep{Ivanova21}. As briefly mentioned in section~\ref{intro}, constraints on the velocities were based on Doppler shifts of absorption lines in high resolution spectra of about 120 targets stars distributed in the volume at known distances. Profile fitting of neutral potassium lines provided between one and four velocity components for each star. The synthesis of the velocities, absorption strengths and distances for the whole set of targets allowed to assign radial velocities to a series of the structures seen in extinction and located along the lines of sight to the target stars. The average uncertainty on the velocity was on the order of 1.0 to 2.0\,{\kms}, as a result of calibration uncertainties and profile fitting simplifications. Note that, due to the limited number of targets, the location of the dust concentration responsible for a given velocity component was rather imprecise, especially in regions of clumpy structures close to each other. 

Figs.~\ref{fig:coupe1}, \ref{fig:coupe2}, \ref{fig:coupe3} and \ref{fig:coupe4} show how our automated velocity assignments compare with these independent, non-automated assignments.  The velocities found by \cite{Ivanova21}  are marked with circles in the figures, at the locations manually assigned by the authors. The color scale used for both types of results on velocities is the same, allowing to compare their values. Again, we emphasize that these absorption-based results are approximate locations. Because they come from line-of-sight integrated data, there is some freedom for the location assignment but there are difficulties in regions where there are several maxima of extinction between the available target stars. Moreover, the locations for the assigned velocities were most of time displaced from the lines of sight to the target stars to avoid overlapping. Finally, our automated method here provides velocities associated with the densest parts of the clouds, while \cite{Ivanova21} measures velocities along the exact sight-lines to the targets, which in general do not cross the densest areas. Still, despite these limitations, it is possible to visualize to which high extinction regions they correspond. A look at the series of vertical planes in the figures shows that at locations where non-automated KI determinations were made, there is a general agreement  between the two types of results, given the uncertainties. This gives us confidence in other assignments in areas where KI results are absent. We also have a few disagreements. A careful look at the results for $l=157.5, 160$ and $180\fdeg$  in the Taurus area ($r=130-180$\,pc) reveals that a few (resp. 1, 2 and three) circles are drawn with colors indicating low velocities between -1.5 and +1\,{\kms}, while there are no corresponding crosses with the same color. While for some of the stars the absorbing columns are very small and the detection may be uncertain, for others like HD 285016 (star 41, v= -1.5\,{\kms} ) and HD 284938  (star 44, v= -1\,{\kms}) the absorption is clearly defined (see Figs. 10 and 12 of \cite{Ivanova21}). Indeed, the maps suggest the existence of two structures in main Taurus distant by $\simeq$ 40\,pc, and the discrepant circles were all located close to the more distant one.  The cloud  numbered 52 (see Table \ref{tab:results}) in the external part of Taurus at 167\,pc and centered at $l=178\fdeg$ is the only one in Taurus to be assigned a low velocity of $2.4\pm 1.0$\,{\kms} and may correspond to this low velocity detected with the potassium lines, and the more distant part of Taurus. However, in Fig.~\ref{fig:coupe4} (top) it appears  co-located with the higher velocity clouds, it is given a low quality index, and the velocity of 2.4\,{\kms} is above the one measured with absorption lines. We believe that the difficulty here resides in the insufficient resolution of the extinction map and the resulting lack of disentangling of the two closer and more distant structures. This problem is affecting the particularly complex and wide Taurus area. Except for this discrepancy, there are validations for each clump for which there are values coming from the absorption if our quality index is good (green contours in Figs. ~\ref{fig:coupe1}, \ref{fig:coupe2}, \ref{fig:coupe3} and \ref{fig:coupe4}). For clouds with lower quality index, and apart from the above mentioned region, most of them are in rather good agreement too. 

\subsection{Other comparisons}

\begin{figure}
    \centering
    \includegraphics[width=1\columnwidth]{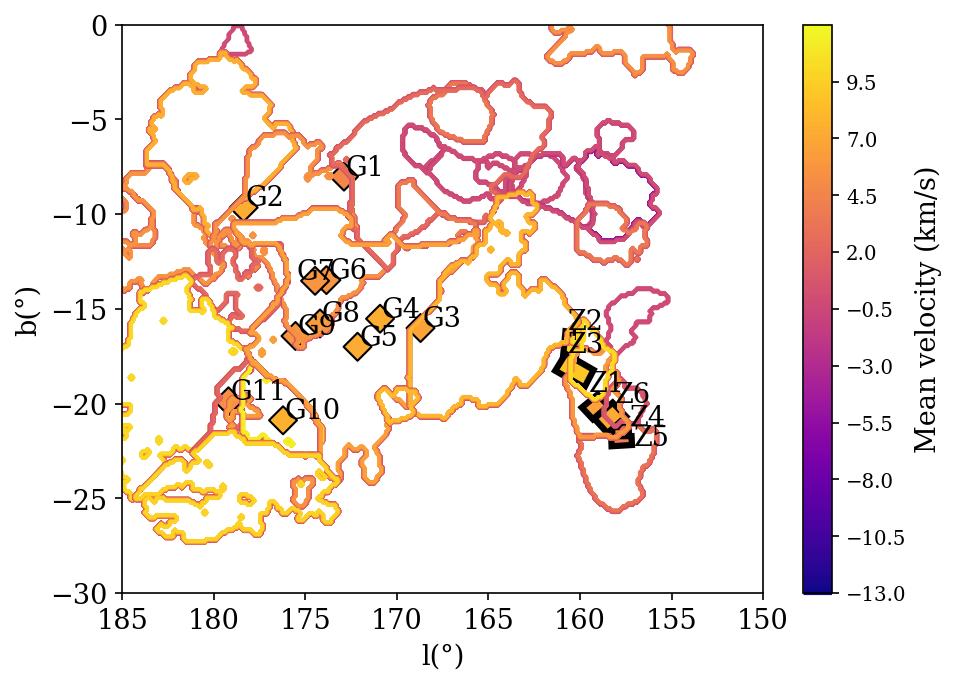}
    \caption{Projection of clumps' contours, colored according to their assigned mean velocity (right scale). The map is in Galactic coordinates. \cite{Zucker18} and \cite{Galli19}'s results are superimposed as markers at their average direction and with the same color scale for their peak-reddening velocities. \cite{Zucker18}'s areas are designated by a Zx annotation, while \cite{Galli19}'s points are marked out by a Gx.}
    \label{fig:comparaison_vit_ZG}
\end{figure}

In Fig.~\ref{fig:comparaison_vit_ZG} and Tables \ref{tab:galli} and \ref{tab:zucker}, we compare our results with those of \cite{Zucker18} and \cite{Galli19} for areas in common. As detailed in section~\ref{intro}, \cite{Zucker18} searched for correspondences between CO emission velocity components and raising steps in cumulative reddening toward known dense structures of Perseus, while \cite{Galli19} derived velocities of a number of Taurus clouds based on the velocities of the young stellar objects from clusters embedded in the clouds. Both studies are reaching a degree of precision superior to our present results. In the case of \cite{Zucker18}, the authors used CO data dedicated to the Perseus area, characterized by very high signal and spectral resolution. In the case of velocities based on YSOs, the authors could use the clustering to improve the precision on the local velocity. Despite these differences, it is informative to compare our assignments with both types of results, since they are obtained in a totally independent way.  Typical uncertainties are less or equal to 0.6\,{\kms} and on the order of 2\,{\kms} for the two works respectively. Fig.~\ref{fig:comparaison_vit_ZG} shows the location of the regions with velocities assigned by both studies, superimposed on  the projected contours of our extinction structures, all colored according to their averaged radial velocity. In the case of the Taurus young star clusters of \cite{Galli19}, they are part of the three distinct series of clouds clearly seen in Figs. \ref{fig:coupe3} and located at increasing distances (about 130, 170 and 200\,pc).
As seen in Tables \ref{tab:galli} and \ref{tab:results}, there is agreement on pairs of distances and velocities given uncertainties and internal velocity dispersion within the clouds, for all points but the clusters G10 (T Tau) and G11 (L1551). These points respectively compare with our extinction regions numbered 15 and 17. Both are part of a very dense region where our decomposition lacks details and where the window used for the velocity assignment is almost fully filled with adjacent clouds, resulting in {\it presence} ratios that are not sufficiently discriminating. For Perseus, the velocities assigned by \cite{Zucker18} seem to agree, within uncertainties, with those of our clumps 173 and 178 (see Table \ref{tab:zucker} and \ref{tab:results}). We note the similarity between the velocity decrease with latitude observed here for longitudes between 155 and 160\fdeg~ and the velocity gradient measured with better precision by \cite{Zucker18}.  


\begin{table*}[b]
    \centering
    \begin{tabular*}{\textwidth}{@{\extracolsep{\fill}} *{7}{c}}
         &   cloudname &     l &     b &  dist &  v$_{lsr}$ & nearby clumps IDs\\
          & & ($^\circ$) & ($^\circ$) & (pc) & {\kms} & \\
        \hline
        G1 &           L\_1517\_1519 & 172.89 &  -8.02 & 158 &   4.69 & 66\\
        G2 &                L 1544 & 178.37 &  -9.65 & 171 &   7.50 & 66\\
        G3 &                L 1495 & 168.71 & -16.01 & 130 &   6.74 & 8\\
        G4 &             B 213\_216 & 170.92 & -15.52 & 161 &   7.49 & 9\\
        G5 &                 B 215 & 172.15 & -17.00 & 129 &   7.38 & 9\\
        G6 &        Heiles Cloud 2 & 173.85 & -13.48 & 141 &   5.75 & 9, 66\\
        G7 &        Heiles Cloud 2 & 174.47 & -13.52 & 136 &   5.78 &9, 66\\
        G8 & L 1535\_1529\_1531\_1524 & 174.20 & -15.77 & 130 &   6.31 & 9, 66 \\
        G9 &                L 1536 & 175.54 & -16.43 & 161 &   5.75 & 9 \\
        G10 &           T Tau cloud & 176.22 & -20.88 & 146 &   7.75 & 15\\
        G11 &                L 1551 & 179.20 & -19.88 & 145 &   5.41 & 17\\
    \end{tabular*}
    \caption{Table of position and velocities for points coming from \cite{Galli19}.}
    \label{tab:galli}
\end{table*}

\begin{table*}[b]
    \centering
    \begin{tabular*}{\textwidth}{@{\extracolsep{\fill}} *{7}{c}}
         &   cloudname &     mean l &     mean b &  dist &  v$_{lsr}$ & nearby clumps IDs\\
          & & ($^\circ$) & ($^\circ$) & (pc) & {\kms}& \\
        \hline
       Z1 &        B1 &  159.1416 & -20.03816 &  294 &   6.7 & 173\\
       Z2 &        B5 &  160.5564 &  -16.7217 &  297 &   9.7 & 173\\
       Z3 &     IC348 &   160.386 & -18.07402 &  289 &   9.0 & 173\\
       Z4 &     L1448 &  157.6976 & -21.41834 &  281 &   4.8 & 173, 178\\
       Z5 &     L1451 &   157.618 &  -21.9173 &  287 &   4.2 & 173, 178\\
       Z6 &   NGC1333 &  158.3264 & -20.58454 &  296 &   7.3 & 173, 178\\
    \end{tabular*}
    \caption{Table of position and velocities for areas coming from \cite{Zucker18}.}
    \label{tab:zucker}
\end{table*}

\begin{table*}[b]
    \begin{floatrow}[3]
        \centering
        {\begin{tabular*}{\columnwidth/3}{@{\extracolsep{\fill}} *{4}{c}|}
        l &     b &  dist &  v$_{lsr}$\\
        ($^\circ$) & ($^\circ$) & (pc) & {\kms} \\
        \hline
           155.0 &           -4.2 &      135 &             5.0 \\
           155.0 &          -10.5 &      274 &             3.0 \\
           155.0 &          -11.2 &      310 &            -1.1 \\
           155.0 &          -14.5 &      152 &             1.9 \\
           157.5 &           -5.3 &      561 &           -16.3 \\
           157.5 &           -5.3 &      453 &            -6.2 \\
           157.5 &           -4.5 &      256 &            -2.0 \\
           157.5 &           -4.3 &      135 &             3.7 \\
           157.5 &           -5.5 &      566 &           -22.0 \\
           157.5 &           -5.7 &      451 &            -9.7 \\
           157.5 &           -5.6 &      256 &            -1.8 \\
           157.5 &           -5.7 &      135 &             4.7 \\
           157.5 &           -9.8 &      147 &             3.3 \\
           157.5 &          -10.7 &      305 &             5.1 \\
           157.5 &          -10.4 &      438 &            -8.1 \\
           157.5 &          -10.0 &      258 &             1.4 \\
           157.5 &          -10.5 &      401 &            -4.1 \\
           157.5 &          -10.9 &      259 &             0.5 \\
           157.5 &          -11.6 &      285 &             3.9 \\
           157.5 &          -11.4 &      397 &            -4.2 \\
           157.5 &          -12.2 &      306 &             2.1 \\
           157.5 &          -12.2 &      400 &            -4.9 \\
           157.5 &          -14.5 &      149 &             1.9 \\
           157.5 &          -23.4 &      157 &             4.2 \\
           157.5 &          -25.0 &      158 &             4.3 \\
           157.5 &          -20.8 &      277 &             5.0 \\
           160.0 &          -34.4 &      142 &             6.0 \\
           160.0 &          -12.2 &      143 &             8.3 \\
           160.0 &          -18.0 &      284 &             8.1 \\
           160.0 &          -17.4 &      291 &             7.5 \\
           160.0 &          -16.8 &      287 &             8.1 \\
           160.0 &          -19.0 &      286 &             7.7 \\
           160.0 &          -18.4 &      285 &             7.8 \\
           160.0 &          -13.0 &      292 &             5.1 \\
           160.0 &           -4.4 &      260 &            -1.4 \\
           160.0 &           -5.8 &      256 &             2.7 \\
           160.0 &          -14.9 &      155 &             1.3 \\
           160.0 &          -11.2 &      295 &             3.9 \\
           160.0 &          -14.5 &      139 &             3.5 \\
           160.0 &          -16.7 &      136 &             2.7 \\
           160.0 &          -21.3 &      278 &             3.3 \\
           160.0 &          -21.8 &      278 &             4.2 \\
           160.0 &          -22.5 &      282 &             3.1 \\
           160.0 &          -11.2 &      462 &            -4.2 \\
           160.0 &           -5.2 &      431 &            -2.6 \\
           160.0 &           -6.0 &      477 &            -8.5 \\
           160.0 &           -4.8 &      465 &            -7.4 \\
           160.0 &           -4.7 &      507 &           -12.8 \\
           160.0 &          -19.6 &      143 &             4.6 \\
           160.0 &          -18.8 &      143 &             7.2 \\
           160.0 &          -17.3 &      283 &             6.0 \\
           160.0 &          -13.4 &      142 &             7.9 \\
           160.0 &           -5.3 &      261 &             1.4 \\
           160.0 &           -5.6 &      441 &            -2.7 \\

    \end{tabular*}}
        {\begin{tabular*}{\columnwidth/3}{|@{\extracolsep{\fill}} *{4}{c}|}
        l &     b &  dist &  v$_{lsr}$\\
        ($^\circ$) & ($^\circ$) & (pc) & {\kms} \\
        \hline
           160.0 &          -17.6 &      139 &             4.4 \\
           160.0 &          -15.7 &      140 &             1.4 \\
           162.5 &          -34.4 &      158 &             8.8 \\
           162.5 &           -8.4 &      151 &             3.8 \\
           162.5 &          -17.3 &      145 &             6.4 \\
           162.5 &          -15.2 &      147 &             6.2 \\
           162.5 &          -16.3 &      147 &             6.1 \\
           162.5 &           -4.2 &      175 &             2.1 \\
           162.5 &           -5.6 &      173 &             2.8 \\
           162.5 &           -6.5 &      177 &             2.1 \\
           162.5 &           -5.6 &      431 &            -5.5 \\
           162.5 &           -4.6 &      495 &            -9.7 \\
           162.5 &           -5.2 &      431 &            -3.5 \\
           162.5 &          -33.4 &      300 &            -2.9 \\
           165.0 &          -24.4 &      136 &             8.0 \\
           165.0 &          -15.9 &      140 &             6.1 \\
           165.0 &           -7.7 &      156 &             7.4 \\
           165.0 &           -9.1 &      159 &             5.3 \\
           165.0 &           -7.7 &      186 &             1.5 \\
           165.0 &           -6.9 &      176 &             3.1 \\
           165.0 &           -5.7 &      180 &             3.2 \\
           165.0 &           -8.7 &      476 &            -1.7 \\
           165.0 &           -7.8 &      478 &            -1.2 \\
           165.0 &           -7.0 &      492 &            -1.0 \\
           167.5 &           -9.7 &      107 &             3.9 \\
           167.5 &          -11.4 &      163 &             2.7 \\
           167.5 &          -11.8 &      138 &             5.9 \\
           167.5 &          -14.1 &      128 &             5.6 \\
           167.5 &          -15.5 &      136 &             6.6 \\
           167.5 &          -17.3 &      136 &             5.9 \\
           167.5 &          -22.3 &      134 &             7.3 \\
           167.5 &          -25.3 &      132 &             7.3 \\
           167.5 &          -26.4 &      138 &             7.5 \\
           167.5 &           -4.3 &      201 &             2.8 \\
           167.5 &          -24.1 &      130 &             7.3 \\
           170.0 &          -26.4 &      138 &             7.7 \\
           170.0 &          -17.3 &      136 &             7.4 \\
           170.0 &          -14.4 &      142 &             6.6 \\
           170.0 &          -13.1 &      146 &             5.4 \\
           170.0 &          -12.2 &      143 &             6.6 \\
           170.0 &          -12.6 &      175 &             2.5 \\
           170.0 &           -4.5 &      190 &             2.5 \\
           170.0 &          -16.7 &      160 &             3.0 \\
           172.5 &           -3.1 &      370 &            -7.5 \\
           172.5 &          -37.2 &      250 &            -6.9 \\
           172.5 &           -3.7 &      230 &            -1.9 \\
           172.5 &           -9.8 &      183 &             4.4 \\
           172.5 &          -15.0 &      129 &             5.8 \\
           172.5 &          -13.3 &      140 &             6.5 \\
           172.5 &          -16.8 &      133 &             7.2 \\
           172.5 &          -16.6 &      159 &             3.6 \\
           172.5 &          -17.8 &      130 &             8.5 \\
           172.5 &          -21.0 &      133 &             8.8 \\
           172.5 &          -10.7 &      143 &             6.3 \\
 
    \end{tabular*}}
        {\begin{tabular*}{\columnwidth/3}{|@{\extracolsep{\fill}} *{4}{c}}
        l &     b &  dist &  v$_{lsr}$\\
        ($^\circ$) & ($^\circ$) & (pc) & {\kms} \\
        \hline
           172.5 &           -9.8 &      142 &             6.1 \\
           172.5 &           -6.7 &      162 &             7.1 \\
           172.5 &           -6.0 &      190 &             5.9 \\
           172.5 &           -4.4 &      195 &             2.5 \\
           172.5 &          -37.9 &      179 &             4.9 \\
           172.5 &          -19.3 &      129 &             8.0 \\
           172.5 &          -19.4 &      153 &             3.9 \\
           172.5 &          -18.4 &      154 &             3.8 \\
           175.0 &           -1.5 &      369 &            -3.2 \\
           175.0 &          -17.3 &      163 &             4.9 \\
           175.0 &          -23.7 &      135 &            11.2 \\
           175.0 &          -17.5 &      132 &             8.8 \\
           175.0 &          -14.9 &      137 &             8.0 \\
           175.0 &          -11.8 &      147 &             3.0 \\
           175.0 &          -20.2 &      126 &            12.0 \\
           175.0 &          -21.2 &      157 &             7.0 \\
           175.0 &           -1.2 &      234 &            -3.2 \\
           177.5 &          -10.8 &      165 &             6.3 \\
           177.5 &          -30.8 &      151 &             8.6 \\
           177.5 &          -29.1 &      132 &             8.9 \\
           177.5 &           -7.8 &      192 &             9.2 \\
           177.5 &           -1.7 &      228 &             2.9 \\
           177.5 &          -11.5 &      196 &            -1.0 \\
           180.0 &          -20.0 &      148 &             8.4 \\
           180.0 &          -17.7 &      153 &             8.3 \\
           180.0 &          -14.2 &      165 &             9.7 \\
           180.0 &          -15.0 &      156 &             8.0 \\
           180.0 &          -16.3 &      155 &             5.6 \\
           180.0 &          -12.9 &      175 &             7.4 \\
           180.0 &          -10.4 &      183 &             7.1 \\
           180.0 &           -7.7 &      186 &             8.2 \\
           180.0 &          -11.4 &      209 &             2.9 \\
           180.0 &          -12.8 &      204 &             2.9 \\
           180.0 &          -15.9 &      202 &            -1.5 \\
           180.0 &          -13.5 &      215 &             0.8 \\
           180.0 &          -11.9 &      219 &            -1.0 \\
           180.0 &           -7.1 &      201 &             4.6 \\
           180.0 &           -8.9 &      207 &             3.6 \\
           180.0 &           -2.7 &      190 &             5.4 \\
           180.0 &           -2.0 &      198 &             6.5 \\
           180.0 &           -1.0 &      222 &             3.0 \\
           180.0 &           -2.5 &      229 &            -1.0 \\
           180.0 &          -34.8 &      155 &             9.9 \\
           180.0 &          -11.5 &      176 &             6.3 \\
           180.0 &          -12.4 &      215 &            -0.9 \\
           180.0 &          -17.2 &      184 &             3.3 \\
           180.0 &          -19.5 &      176 &             4.1 \\
           182.5 &          -34.8 &      155 &             8.5 \\
           182.5 &           -4.0 &      215 &             5.7 \\
           182.5 &           -7.9 &      196 &             8.0 \\
           182.5 &           -8.9 &      227 &             3.0 \\
           182.5 &          -36.9 &      156 &             9.9 \\
           \\
           \\
    \end{tabular*}}
    \caption{Table of position and velocities for points coming from \cite{Ivanova21} and represented by colored circles in Figs.~\ref{fig:coupe1}, \ref{fig:coupe2}, \ref{fig:coupe3} and \ref{fig:coupe4} }
    \label{tab:potassium1}
    \end{floatrow}
\end{table*}

\begin{table}[b]
    \centering
    \begin{tabular*}{\columnwidth}{@{\extracolsep{\fill}} *{4}{c}}
        l &     b &  dist &  v$_{rad}$ \\
        ($^\circ$) & ($^\circ$) & (pc) & {\kms} \\
        \hline
    153.875 &  -1.625 &  135.0 &  5.7 \\
     153.875 &    -1.5 &  135.0 &  5.6 \\
       154.0 &    -2.0 &  136.0 &  5.4 \\
       154.0 &  -1.875 &  137.0 &  5.6 \\
       154.0 &   -1.75 &  137.0 &  5.6 \\
       154.0 &  -1.625 &  137.0 &  5.6 \\
       154.0 &    -1.5 &  137.0 &  5.4 \\
       154.0 &  -1.375 &  135.0 &  5.4 \\
     154.125 &   -2.25 &  135.0 &  4.9 \\
     154.125 &  -2.125 &  137.0 &  5.3 \\
     154.125 &    -2.0 &  137.0 &  5.5 \\
     154.125 &  -1.875 &  137.0 &  5.5 \\
     154.125 &   -1.75 &  137.0 &  5.6 \\
     154.125 &  -1.625 &  137.0 &  5.4 \\
     154.125 &    -1.5 &  137.0 &  5.4 \\
     154.125 &  -1.375 &  137.0 &  5.3 \\
      154.25 &  -2.375 &  137.0 &  5.3 \\
      154.25 &   -2.25 &  137.0 &  4.9 \\
      154.25 &  -2.125 &  137.0 &  5.4 \\
      154.25 &    -2.0 &  137.0 &  5.5 \\
      154.25 &  -1.875 &  137.0 &  5.6 \\
      154.25 &   -1.75 &  137.0 &  5.4 \\
      154.25 &  -1.625 &  137.0 &  5.2 \\
      154.25 &    -1.5 &  137.0 &  5.3 \\
      154.25 &  -1.375 &  137.0 &  5.2 \\
     154.375 &    -2.5 &  137.0 &  5.3 \\
     154.375 &  -2.375 &  137.0 &  5.5 \\
     154.375 &   -2.25 &  137.0 &  5.4 \\
     154.375 &  -2.125 &  137.0 &  5.6 \\
     154.375 &    -2.0 &  137.0 &  5.6 \\
     154.375 &  -1.875 &  137.0 &  5.5 \\
     154.375 &   -1.75 &  137.0 &  5.3 \\
     154.375 &  -1.625 &  137.0 &  5.2 \\
     154.375 &    -1.5 &  137.0 &  5.2 \\
     154.375 &  -1.375 &  137.0 &  5.2 \\
       154.5 &    -2.5 &  137.0 &  5.2 \\
       154.5 &  -2.375 &  137.0 &  5.3 \\
       154.5 &   -2.25 &  137.0 &  5.5 \\
       154.5 &  -2.125 &  137.0 &  5.7 \\
       154.5 &    -2.0 &  137.0 &  5.6 \\
       154.5 &  -1.875 &  137.0 &  5.5 \\
       154.5 &   -1.75 &  137.0 &  5.5 \\
       154.5 &  -1.625 &  137.0 &  5.1 \\
       154.5 &    -1.5 &  137.0 &  5.1 \\
       154.5 &  -1.375 &  135.0 &  5.1 \\
     154.625 &    -2.5 &  137.0 &  5.4 \\
     154.625 &  -2.375 &  137.0 &  5.5 \\
     154.625 &   -2.25 &  137.0 &  5.9 \\
     154.625 &  -2.125 &  137.0 &  6.0 \\
     154.625 &    -2.0 &  137.0 &  5.8 \\
     154.625 &  -1.875 &  137.0 &  5.5 \\
     154.625 &   -1.75 &  137.0 &  5.3 \\
     154.625 &  -1.625 &  137.0 &  5.3 \\
     154.625 &    -1.5 &  137.0 &  5.2 \\
     154.625 &  -1.375 &  135.0 &  5.1 \\
      154.75 &  -2.375 &  137.0 &  5.7 \\
      154.75 &   -2.25 &  137.0 &  5.9 \\
      154.75 &  -2.125 &  137.0 &  6.0 \\

    \end{tabular*}
    \caption{Coordinates and radial velocities for the grid of points shown in Figs.~\ref{fig:coupe1}, \ref{fig:coupe2}, \ref{fig:coupe3} and \ref{fig:coupe4} (first lines). The entire table is available online at the CDS. The typical uncertainty in velocity is 1.3\,{\kms}.}
    \label{tab:points}
\end{table}

\section{Conclusions and perspectives}\label{perspectives}

We have presented a novel technique of automated kinetic tomography of dense molecular clouds. The method is based on 3D dust extinction density maps and CO emission spectra, and the assignment of radial velocities to the clouds is uniquely based on morphological criteria, namely properties of projected contours of extinction structures on the one hand and on contours associated to CO emission in specific velocity intervals on the other hand. Individual extinction structures are issued from a Fellwalker discretization of the 3D dust distribution. The method was tested on the set of nearby clouds forming the Taurus, Auriga, Perseus and California groups. The choice of this particularly complex CO-rich area is motivated by the availability of comparisons with previous, independent results. The application of the technique allowed to assign a velocity to a large fraction of the dust structures issued from Fellwalker (42 from a total of 45). The three remaining structures have direction and distance very similar to neighboring structures with assigned velocities, and they may be simply continuations of those neighbors. We ranked the assignments based on geometrical criteria. Except for a few low confidence  associations, we found good agreement with independent determinations of velocities, whatever their sources, synthesis of KI absorption towards individual stars, combination of reddening measurements along dense clouds directions and corresponding CO spectra, or YSO clustering in the direction of known molecular clouds. 

Our results show that an automated kinetic tomography based on morphological criteria and applied to structures extracted from 3D extinction density distribution on the one hand, and radio emission data on the other hand, is a realistic objective. They demonstrate the rare occurrence of situations with two or more distinct clouds at different velocities, distributed in distance along the same direction, and with projected contours so similar that disentangling is precluded.  The results suggest that the application of the method to larger volumes and larger velocity differences between the groups of clouds is feasible. Work is in progress in this direction. 

The application of the technique to other volumes of gas and dust will necessitate to adapt the values of the various parameters entering the algorithm, and in particular the number of velocities treated separately. Their optimal choice depends on the dust map spatial resolution and the number of discrete structures issued from the decomposition of the continuous extinction density distribution, on the spectral resolution of the emission data and the noise level, and, finally, on the extent of the region of space to be treated. In particular, the Fellwalker decomposition has some arbitrary aspects and other types of decomposition should be attempted. This is beyond the scope of this work whose main aim is to validate a technique. In future, one may expect more resolved and more extended 3D extinction maps, due to the continuous production of massive amounts of parallaxes and photometric data from Gaia, as well as data from ground-based surveys. This should allow to reach more details and take better advantage of the radio spectral resolution or of more sensitive radio data.  Although the application of such a technique to large distances and low latitudes, that is to a large number of clouds with overlapping projections, is expected to be increasingly difficult, it should be facilitated by the fact that the projections of the distant clouds are in general smaller than those of the foreground objects, and as a consequence the contours should be disentangled thanks to the ratios discussed above. 

It remains that some structures lacked velocity assignments, mainly because 3D extinction maps are of limited resolution and are not realistic enough in some areas. This is why it is clear that combining or complementing the present technique solely based of morphology with more quantitative arguments, brightness of emission, opacity, absorption lines, is expected to produce better results and certainly deserves further developments. In such combinations the application of the present method would provide a useful first order solution of kinetic tomography. In particular, it could serve as a starting point to the application, region by region, of techniques similar to the one developed by \cite{Zucker18} that require some preliminary constraints on velocity intervals. 

The application to other tracers (HI, CI$^{+}$) detected in emission is also in principle feasible, and in this case the restriction mask would have to be modified or removed. In the case of HI, the significantly larger width of velocity components associated with the atomic phase would render the analysis significantly more difficult. A preliminary application to CO, and subsequent application to HI with some imposed parameters may be a solution to attempt.

\begin{acknowledgements}

We are grateful to our referee for their quick reading and thoughtful, constructive comments and criticisms.\\
Q. D. and C. H. acknowledge the CNES Gaia support, 
with a funding under ``identifiant action 6187''. 
 J.-L. V.  acknowledges support from the EXPLORE project. EXPLORE has received funding from the European Union’s Horizon 2020 research and innovation programme under grant agreement No 101004214.

This work has made use of data from the European Space Agency (ESA) mission Gaia (https://www.cosmos.esa.int/gaia), processed by the Gaia Data Processing and Analysis Consortium (DPAC, https://www.cosmos.esa.int/web/gaia/dpac/consortium). Funding for the DPAC has been provided by national institutions, in particular the institutions participating in the Gaia Multilateral
Agreement. This work also makes use of
data products from the 2MASS, which is a joint project of
the University of Massachusetts and the Infrared Processing
and Analysis Center/California Institute of Technology,
funded by the National Aeronautics and Space Administration
and the National Science Foundation.\\
This research has made use of the SIMBAD database, operated at CDS, Strasbourg, France.
\end{acknowledgements}

%
\bibliographystyle{aa} 
\bibliography{mybib} 
%

\appendix

\section{Restriction mask}\label{restriction_masks}

\begin{figure}
    \centering
    \includegraphics[width=0.8\columnwidth]{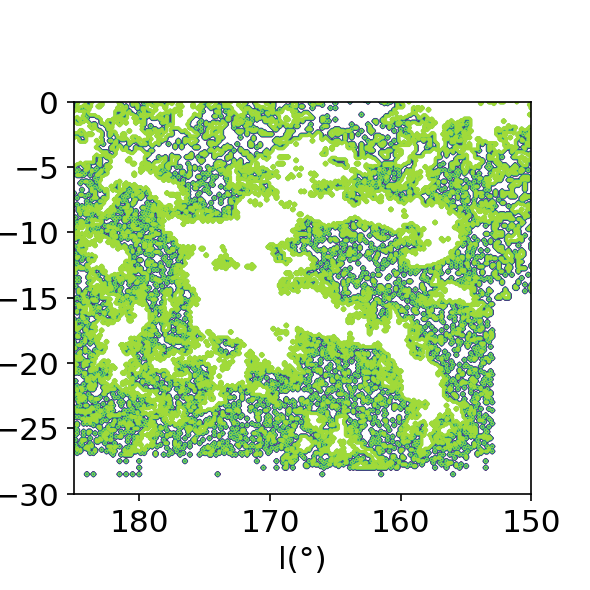}
    \includegraphics[width=0.8\columnwidth]{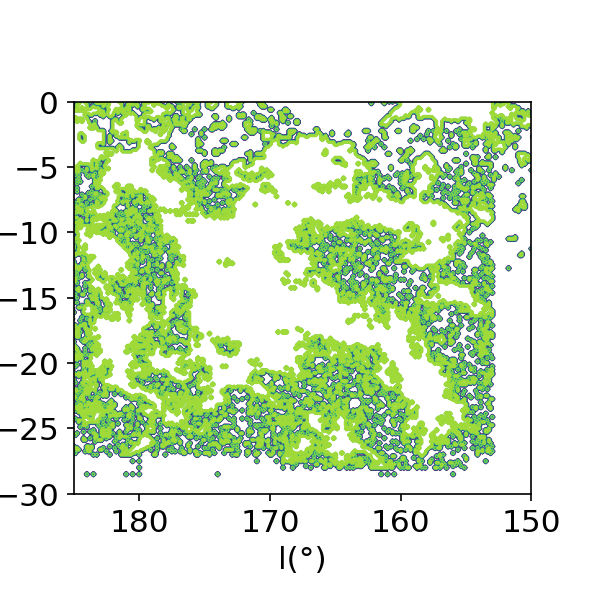}
    \includegraphics[width=0.8\columnwidth]{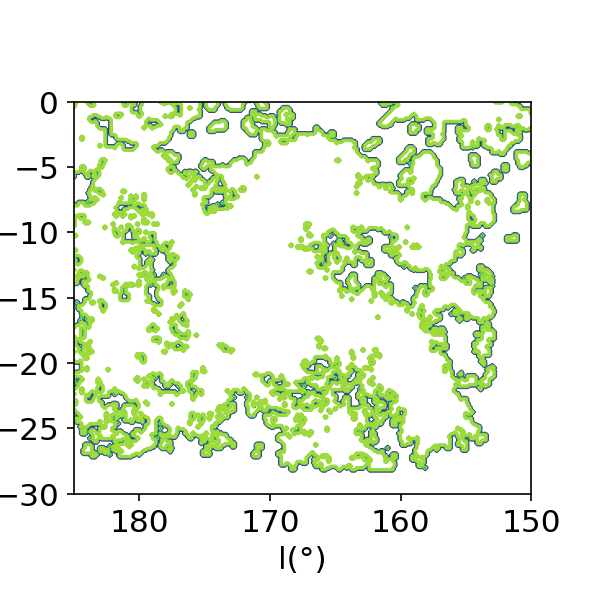}
    \caption{{\bf Top:} Restriction mask obtained by estimating the noise on the velocity-integrated \cite{Dame2001} CO data and keeping the lines where the integrated intensity is above the threshold (0.84K\,{\kms}). {\bf Middle:} Mask obtained after de-noising the CO signal with \ROHSA, then keeping all the lines that have at least a position-position-velocity above the threshold used for Fellwalker, namely 0.20 K. {\bf Bottom:} Mask obtained by applying the Fellwalker algorithm on the post-{\ROHSA} data. This is the mask we used in our search for velocity assignment.}
    \label{fig:restriction_masks}
\end{figure}

The role of the restriction mask mentioned in section \ref{codata}  is to eliminate from the 3D extinction distribution those volumes whose projections on the sky correspond to no CO emission. It must encompass all the features of the CO map, however, one wants to avoid too complex contours. Fig.~\ref{fig:restriction_masks} illustrates three approaches including the selected one. The easiest way would be to integrate the CO data over the full velocity range, to estimate the noise level for this integrated map, and to use it as a threshold value. This leads to a very noisy  mask (shown in Fig.~\ref{fig:restriction_masks}:top), and some features, such as the cloud at $l=155^\circ$ and $b=-15^\circ$, do not appear clearly. We can go one step further by de-noising the CO map using {\tt ROHSA}. It has the effect of maintaining in the exclusion process isolated pixels with weak intensity, corresponding to noisy data. However, there is a risk to also eliminate real weak features characterized by a very narrow velocity range. To avoid the loss of such narrow and weak features of the spectrum, we defined a threshold based on the standard deviation of the individual CO spectral cube elements, instead of using the velocity-integrated elements. This method is slightly better, but far from perfect, as shown in Fig.~ \ref{fig:restriction_masks}:middle. We finally built a mask based on the output of the Fellwalker algorithm applied to the CO map, after preliminary de-noising by means of {\tt ROHSA}. We used the same threshold value as the one used in the previous step (see also Part \ref{codata}). The Fellwalker algorithm allows us to identify actual structures out of the noise. The mask based on the third method is also shown in Fig.~\ref{fig:contours_clumps}.

\end{document}